\documentclass{article}

\usepackage{microtype}
\usepackage{graphicx}
\providecommand{\seqsplit}[1]{\texttt{\sloppy\hspace{0pt}#1}}
\usepackage{booktabs}
\usepackage{multirow}
\usepackage{float}
\usepackage{xcolor}
\usepackage{tikz}
\usetikzlibrary{arrows.meta,positioning,fit,backgrounds,calc}
\PassOptionsToPackage{hyphens}{url}
\usepackage{hyperref}
\usepackage[accepted]{icml2026}
\makeatletter
\renewcommand{\Notice@String}{}
\makeatother
\hypersetup{pdfsubject={ICML 2026 FoGen Workshop}}

\usepackage{amsmath}
\usepackage{amssymb}
\usepackage{mathtools}

\usepackage[capitalize,noabbrev]{cleveref}
\usepackage{balance}

\icmltitlerunning{Probe choice changes canary-memorization verdicts in a LoRA-tuned autoregressive testbed}

\begin{document}

\twocolumn[
\icmltitle{Probe Choice Changes Canary-Memorization Verdicts:\\
Three Post-Hoc Disagreement Case Studies in a\\
Text-Dominant LoRA-Tuned Autoregressive Testbed}

\begin{icmlauthorlist}
\icmlauthor{Zhichao Fan}{uiuc}
\icmlauthor{Zexin Zhuang}{smu}
\icmlauthor{Yanhang Li}{northeastern}
\end{icmlauthorlist}

\icmlaffiliation{uiuc}{University of Illinois Urbana-Champaign}
\icmlaffiliation{smu}{Southern Methodist University}
\icmlaffiliation{northeastern}{Northeastern University}

\icmlcorrespondingauthor{Zhichao Fan}{zhichao8@illinois.edu}
\icmlcorrespondingauthor{Zexin Zhuang}{zexinz@smu.edu}
\icmlcorrespondingauthor{Yanhang Li}{li.yanha@northeastern.edu}

\icmlkeywords{memorization, generalization, autoregressive generative models, evaluation metrics, canary probes, foundations of generative models}

\vskip 0.3in
]

\printAffiliationsAndNotice{}

\begin{abstract}
We audit a fixed prefix-window mean-NLL memorization probe
($K\!=\!20$) on a Qwen2.5-VL-7B canary testbed and report three
post-hoc cases where it disagrees with full-span secret NLL or
greedy exact-recall.
\textbf{C3 (false negative, window truncation):} damage lands on
hex tokens \emph{outside} $K\!=\!20$; the probe stays flat while
hit$@1$ drops.
\textbf{C4 (false positive, non-secret drift):} the probe moves,
but $\sim\!99\%$ sits on non-secret preamble; the secret span and
hit$@1$ are unchanged.
\textbf{C5 (ambiguous in-window drop):} the probe falls on an
undertrained baseline while full-span hex is positive and
hit$@1\!=\!0$.
Recommendation: report (i) full-span secret NLL, (ii) a
span-localised decomposition, (iii) behavioural exact-recall at
$k\!\geq\!4$, and (iv) decoy probes before asserting
secret-specificity. Evidence is on controlled canaries in one
backbone; magnitudes are testbed-specific.
\end{abstract}

\section{Introduction}
\label{sec:intro}

A foundational question across machine learning is whether observed
behaviour reflects \emph{generalization} from training data or
\emph{memorization} of specific training points \citep{feldman2020does}.
For deep generative models, this question has been studied through
canaries, extraction, and dataset deduplication
\citep{carlini2019secret,carlini2021extracting,carlini2023quantifying,lee2022deduplicating}.
Operationalising it requires probes that map a model's distribution
over outputs to a numeric memorization signal. In the autoregressive
setting --- which covers language models and the language tower of
vision-language and multimodal generative models --- three probe
families dominate:
(a) per-token \emph{negative log-likelihood} (NLL) averaged over a
fixed token window (often the first $K$ tokens of a target),
(b) \emph{full-span} secret NLL on the substring whose memorization
is claimed, and
(c) \emph{behavioral exact-recall}, i.e.\ whether greedy or sampled
decoding emits the secret string.
Closely related quantities (loss thresholding, perplexity ratios
against a held-out reference, calibrated likelihood ratios) appear
in adjacent memorization-evaluation literature.

\textbf{Why probe choice matters.}
A central concern motivating this paper's audit is that probe choice
can change the false-positive and false-negative rates of a
memorization claim --- a probe-versus-probe consistency question that
is separable from the underlying memorization rate. Claims about
whether a generative model has memorized rest on which probe was used
to call the verdict, and audits asking ``does the standard probe
agree with itself across natural alternatives?'' are a
template-specific probe-audit question --- one that can confound
broader memorization-vs-generalization conclusions if a single window
is used uncritically. In related settings, \citet{liu2026whack} show
that benign or task-style fine-tuning can activate verbatim recall in
a generation-based extraction setting, while \citet{borkar2025ripple}
show privacy-ripple effects from adding or removing personal
information during language-model training. We do not refute or
replicate those reactivation findings, but use a related
memorization-reactivation probe setting on a different testbed and
document where it is and is not self-consistent.

\textbf{Our testbed.}
We use a single autoregressive generative-model testbed: a Qwen2.5-VL-7B
\citep{bai2025qwen25vl} backbone with frozen vision tower and
language-tower LoRA \citep{hu2022lora} into which we inject $20$ image
canaries and $20$ text canaries (text canaries pair with a fixed
$224\!\times\!224$ mid-gray placeholder image, so the strongest evidence
in this paper is essentially text-only). On top of the canary-injected
base we stack a second LoRA performing benign supervised fine-tuning
(bSFT) on one of four data sources, none of which contain the canary
strings (\S\ref{sec:method}).
This setup gives us $34$ aggregate bSFT cells $=\,102$ seed-level runs
($144$ with $42$ matched-norm \textsc{LoRA-Noise} controls), and an
explicit \emph{undertrained-canary} headroom regime that lets us bound
how far reactivation could move the probe in principle (App.~\ref{app:scope}).
Across $102$ bSFT seed-level runs we observe \emph{no} full-secret
NLL recovery and \emph{no} hit$@1$ above baseline
(Tab.~\ref{tab:dissociation}, App.~\ref{app:allcells}); many cells are
saturated at hit$@1\!=\!1.00$ where reactivation cannot be observed
without prior secret damage, so this null is the \emph{condition under
which the probe-disagreement audit is run}, not a substantive null on
benign-SFT reactivation in autoregressive generative models. Our
setting is adjacent to but distinct from large-scale extraction
\citep{carlini2021extracting}, dedup-driven memorization
\citep{lee2022deduplicating}, and training-data unlearning
\citep{bourtoule2021unlearning,maini2024tofu}.

\textbf{Terminology.} We use the following labels throughout:
\begingroup
\raggedright
\begin{itemize}\setlength{\itemsep}{1pt}\setlength{\parskip}{0pt}
\item \emph{full-secret recovery}: sign-consistent negative
$\Delta\mathrm{NLL}_{\mathrm{hex}}$ on the full $13$-token secret span
across seeds, or hit$@1$ above baseline.
\item \emph{undertrained in-window hex-token NLL improvement}:
sign-consistent negative $\Delta\mathrm{NLL}_{\mathrm{mean20}}$ whose
per-span decomposition localises onto $\sim\!9$ in-window hex tokens
(positions $11$--$19$) plus the leading \texttt{canary\_lit}, with
full-span hex $\geq\!0$ and hit$@1$ unchanged.
\item \emph{behavioral damage}: secret-span $\Delta\mathrm{NLL}$
increase \emph{plus} hit$@1$ drop.
\item \emph{non-secret preamble drift}: mean$_{20}$ moves but
$\geq\!90\%$ of the per-span contribution sits on \texttt{preamble},
$\Delta\mathrm{NLL}_{\mathrm{hex}}\!<\!0.5\,|\Delta\mathrm{NLL}_{\mathrm{mean20}}|$,
and hit$@1$ unchanged.
\item \emph{window-truncation failure}: mean$_{20}$ flat while damage
is visible to wider-window probes or hit$@1$.
\end{itemize}
\endgroup

\begin{figure*}[!t]
\centering
\includegraphics[width=0.72\textwidth]{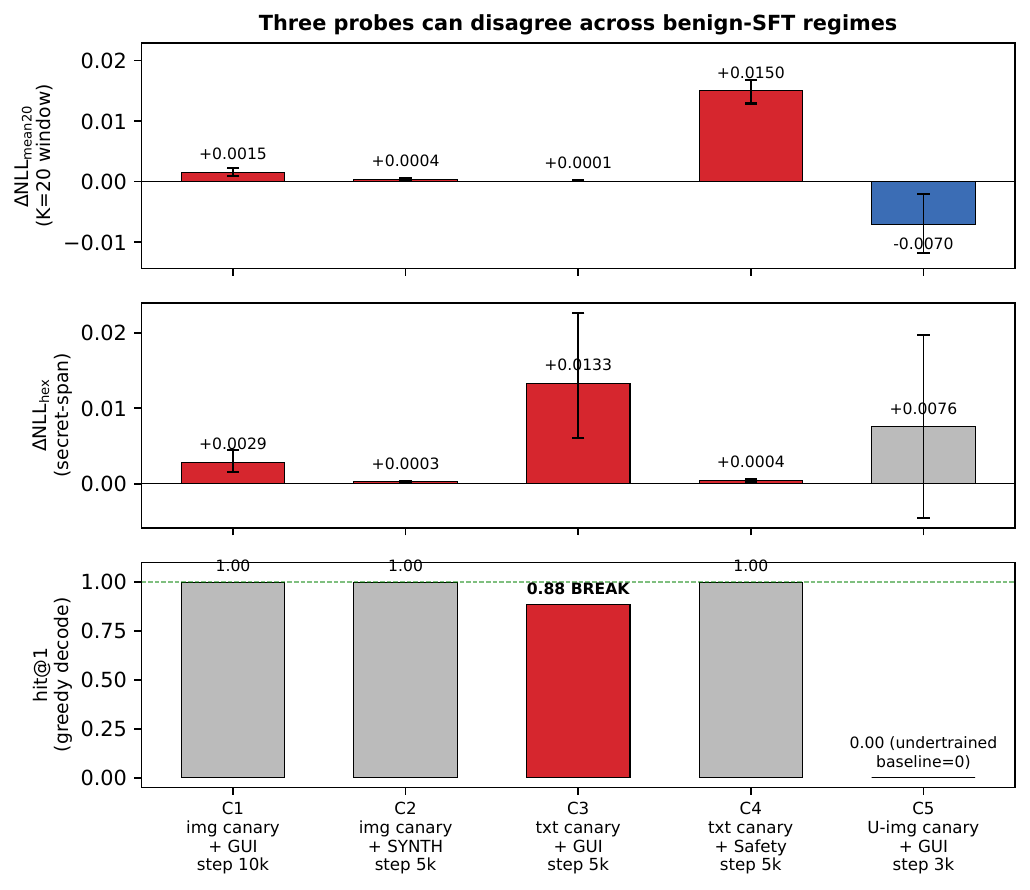}
\caption{\textbf{Three probes can disagree across benign-SFT regimes}
(Qwen2.5-VL-7B, stacked LoRA bSFT). C1: img$+$\textsc{GUI} 10k;
C2: img$+$\textsc{SYNTH} 5k; C3: txt$+$\textsc{GUI} 5k;
C4: txt$+$\textsc{Safety} 5k; C5: U-img$+$\textsc{GUI} 3k.
Top: $\Delta_{\mathrm{mean20}}$ (internally pre-specified probe);
middle: $\Delta_{\mathrm{hex}}$ (full $13$-token secret span);
bottom: greedy hit$@1$.
Error bars are descriptive hierarchical resampling intervals
(outer 3 seeds, inner 20 canaries, $B\!=\!10{,}000$); they are not
population CIs and are not selection-adjusted.
\textbf{Grayscale-readable encoding:} bar \emph{direction} (sign) carries
the qualitative finding; bar \emph{length} carries magnitude; bracket
overlap with the dashed $0$ baseline indicates uncertainty. Colour
(red CI$\!>\!0$, blue CI$\!<\!0$, grey brackets $0$) is redundant.
Only C3 has secret-span NLL and behavioural exact-recall both moving on
narrow tail-token damage; mean$_{20}$ misses it (window truncation).}
\label{fig:dissociation}
\end{figure*}

\textbf{Three probe-disagreement case studies.}
During the same experiments we exhibit three \emph{post-hoc} cases
(Fig.~\ref{fig:dissociation}) where the internally pre-specified
probe $\Delta\mathrm{NLL}_{\mathrm{mean20}}$ gives an answer that
differs from full-span secret NLL or from behavioral exact-recall,
and each case maps to a distinct failure
mode of truncated mean-NLL as an evaluation metric for memorization in
an autoregressive generative model:
\begin{itemize}\setlength{\itemsep}{2pt}\setlength{\parskip}{0pt}
\item \textbf{Case~1 (C3, T-bSFT-GUI 5k --- false negative under window
truncation):} mean$_{20}$ flat ($+0.0001$) while
$\Delta_{\mathrm{hex}}\!=\!+0.0133$ and hit$@1$ drops to $0.88$ ---
damage on a hex-token \emph{outside} $K\!=\!20$.
\item \textbf{Case~2 (C4, T-bSFT-Safety 5k --- false positive from
non-secret span drift):} mean$_{20}\!=\!+0.0150$ sits $\sim\!99\%$ on
non-secret \texttt{preamble}; secret span and hit$@1$ unchanged.
\item \textbf{Case~3 (C5, U-bSFT-GUI 3k --- ambiguous undertrained in-window NLL drop):} mean$_{20}\!=\!-0.0070$ on an undertrained
baseline localises to in-window hex tokens, but the full-span hex
point estimate is positive (and positive across all $3$ seeds; hb
brackets zero) with hit$@1\!=\!0$. \emph{No secret-specificity claim} is
made; format-prior smoothing or template adaptation are equally
consistent without true-vs-decoy probes (a paired decoy audit at the
7B C5 cell remains future work; the smaller-family C5-equivalent
decoy results in Tab.~\ref{tab:decoy} are descriptive stress tests at
\emph{different} cells).
\end{itemize}

\noindent The $K\!=\!20$ probe and the ``no full-secret recovery''
operating condition were chosen before seed-level results were
aggregated, but were \emph{not} preregistered or externally certified;
the three case studies and the selection protocol of
\S\ref{sec:results-failures} are post-hoc.

\textbf{Contributions.}
\begin{itemize}\setlength{\itemsep}{2pt}\setlength{\parskip}{0pt}
\item A template-specific audit of our internally chosen
$K\!=\!20$ truncated mean-NLL probe against full-span secret NLL and
behavioral exact-recall, framed as an evaluation-metric question for
canary-based memorization in autoregressive generative models
(\S\ref{sec:method}).
\item Three post-hoc probe-disagreement case studies illustrating two
mechanical probe mismatches and one unresolved undertrained-control
ambiguity (\S\ref{sec:results-failures}), with full per-cell selection
audit (Tab.~\ref{tab:selection}, App.~\ref{app:selection}) and an
algebraic span decomposition that closes the dissociation to within
bf16 residual (\S\ref{ss:case3}, App.~\ref{app:algebra}).
\item Recommendations for evaluation protocols that can flag the
two failure modes and the ambiguity case before they propagate into
generalization-versus-memorization conclusions
(\S\ref{sec:discussion}, App.~\ref{app:scope}).
\end{itemize}

\textbf{Scope.} Three case studies in one backbone are not a
population claim. We discuss in \S\ref{sec:related}
and \S\ref{sec:discussion} which parts of the failure-mode taxonomy
are likely to transfer to other autoregressive generative models and
which depend on testbed-specific choices (template, tokenizer,
LoRA configuration, frozen vision tower). $n\!=\!3$ outer seeds;
intervals descriptive only; cross-architecture and cross-scale
replication is the natural next experiment.

\section{Method: canaries, bSFT matrix, three probes}
\label{sec:method}

\begin{figure*}[t]
    \centering
    \definecolor{qwenblue}{HTML}{DCEBFA}
\definecolor{canarygreen}{HTML}{DDF4E8}
\definecolor{bsftviolet}{HTML}{E8E2F7}
\definecolor{probea}{HTML}{FFF5E8}
\definecolor{probeb}{HTML}{FFE6C8}
\definecolor{probec}{HTML}{FFD3A3}
\definecolor{groupfill}{HTML}{F7F8FA}
\definecolor{edgegray}{HTML}{4B5563}

\resizebox{0.95\textwidth}{!}{%
\begin{tikzpicture}[
  font=\footnotesize,
  base/.style={draw=blue!60!black, thin, rounded corners=2pt, fill=qwenblue,
    align=center, minimum width=2.75cm, minimum height=3.25cm, inner sep=4pt},
  lora/.style={draw=green!55!black, thin, rounded corners=2pt, fill=canarygreen,
    align=center, minimum width=2.35cm, minimum height=1.55cm, inner sep=4pt},
  bsft/.style={draw=violet!65!black, thin, rounded corners=2pt, fill=bsftviolet,
    align=center, minimum width=2.65cm, minimum height=1.65cm, inner sep=4pt},
  probe/.style={draw=orange!80!black, thin, rounded corners=2pt,
    align=center, minimum width=2.65cm, minimum height=1.15cm, inner sep=4pt},
  verdict/.style={draw=black!45, thin, rounded corners=2pt, fill=black!6,
    align=center, minimum width=1.35cm, minimum height=3.95cm, inner sep=4pt},
  note/.style={font=\scriptsize, align=center, text=edgegray},
  mono/.style={font=\scriptsize\ttfamily, align=center, text=black!82},
  group/.style={draw=black!25, thin, dashed, rounded corners=3pt, fill=groupfill},
  flow/.style={-{Latex[length=2.1mm,width=1.8mm]}, semithick, draw=edgegray},
  pipe/.style={semithick, draw=edgegray},
  weakflow/.style={-{Latex[length=1.6mm,width=1.4mm]}, dotted, semithick, draw=black!55},
  arlabel/.style={font=\scriptsize\ttfamily, fill=white, inner xsep=2pt, inner ysep=1pt, text=black},
]

\node[base] (model) at (0,0)
  {\textbf{Qwen2.5-VL-7B}\\[2pt]
   \scriptsize frozen vision tower\\[1pt]
   \scriptsize language tower\\[-1pt]
   \scriptsize (trainable)};

\node[lora] (canary) at (3.85,0)
  {\textbf{Canary LoRA}\\[2pt]
   \scriptsize $r=32,\ \alpha=64$};

\node[bsft] (bsft) at (8.55,0)
  {\textbf{bSFT LoRA}\\[2pt]
   \scriptsize $r=16,\ \alpha=32,\ \mathrm{lr}=2e{-}5$\\[-1pt]
   \scriptsize $5000$ examples, $n=3$ seeds};

\node[mono] (cdata) at (3.85,2.05)
  {$20$ image canaries\\
   $20$ text canaries\\
   \texttt{CANARY-\{16hex\}-END}};
\node[mono] (bdata) at (8.55,2.05)
  {\textsc{GUI}\\
   \textsc{Synth}\\
   \textsc{Safety}\\
   \textsc{Random}\\[-1pt]
   \normalfont\scriptsize(no canary strings)};

\node[probe, fill=probea] (mean) at (12.75,2.0)
  {\textbf{$\mathrm{mean}_{20}$ probe}\\[1pt]
   \scriptsize $K=20$ prefix window\\[-1pt]
   \scriptsize internally pre-specified};
\node[probe, fill=probeb] (hex) at (12.75,0)
  {\textbf{$\Delta\mathrm{NLL}_{\mathrm{hex}}$ probe}\\[1pt]
   \scriptsize full $13$-token secret span\\[-1pt]
   \scriptsize $K=30$ forward pass};
\node[probe, fill=probec] (hit) at (12.75,-2.0)
  {\textbf{hit@1 probe}\\[1pt]
   \scriptsize greedy behavioral recall\\[-1pt]
   \scriptsize $T=0.7,\ p=0.95$};

\node[verdict] (verdict) at (15.75,0)
  {\textbf{verdicts}\\[-1pt]
   \textbf{disagree}\\[4pt]
   \scriptsize C3\\[-1pt]
   \scriptsize C4\\[-1pt]
   \scriptsize C5};

\draw[flow] (model.east) -- node[above, note] {inject} (canary.west);
\draw[flow] (canary.east) -- (bsft.west);
\draw[flow] (cdata.south) -- (canary.north);
\draw[flow] (bdata.south) -- (bsft.north);

\coordinate (split) at (10.55,0);
\coordinate (topbranch) at (10.55,2.0);
\coordinate (botbranch) at (10.55,-2.0);
\draw[pipe] (bsft.east) -- (split);
\draw[pipe] (botbranch) -- (topbranch);
\draw[flow] (topbranch) -- (mean.west);
\draw[flow] (split) -- (hex.west);
\draw[flow] (botbranch) -- (hit.west);

\draw[weakflow] (mean.east) -- (verdict.west |- mean.east);
\draw[weakflow] (hex.east) -- (verdict.west);
\draw[weakflow] (hit.east) -- (verdict.west |- hit.east);

\begin{scope}[on background layer]
  \node[group, fit=(model)(canary)(bsft)(cdata)(bdata), inner sep=9pt] (pipeline) {};
\end{scope}
\node[font=\scriptsize\itshape, text=edgegray, anchor=west]
  at ([xshift=5pt,yshift=-7pt]pipeline.north west)
  {LoRA injection and tuning pipeline};

\end{tikzpicture}}%
    \caption{Single-backbone canary-memorization audit testbed. A canary
    LoRA is merged into Qwen2.5-VL-7B before a canary-string-free bSFT
    LoRA is stacked from one of four data sources. The resulting model is
    read by three independent probes: the internally pre-specified
    $\mathrm{mean}_{20}$ prefix-window NLL probe, a full secret-span
    $\Delta\mathrm{NLL}_{\mathrm{hex}}$ probe, and a behavioral hit@1
    probe. Disagreements among these probes define the C3, C4, and C5
    case studies.}
    \label{fig:method_overview}
\end{figure*}
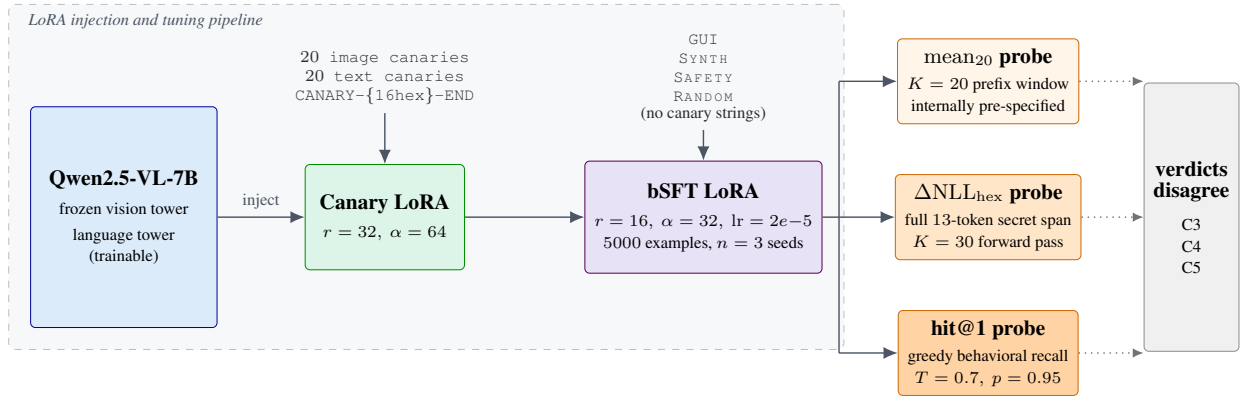

We instantiate a controlled memorization-evaluation testbed
(Fig.~\ref{fig:method_overview}) whose three ingredients --- canary
design, benign supervised fine-tuning (bSFT) matrix, and probe
specification --- jointly determine the generalization-versus-memorization
verdict any given probe will return on this autoregressive generative
model. The choices below
preserve the canary-and-probe setup of \citet{carlini2019secret}
while making the spans and windows on which probes are computed
explicit, so that probe disagreement can be localised to specific
token positions rather than absorbed into a global average.

\paragraph{Canaries.}
Following the small-canary-set tradition of \citet{carlini2019secret},
$N\!=\!20$ canaries per modality (the exact set size is our own
design parameter):
\begin{itemize}\setlength{\itemsep}{1pt}\setlength{\parskip}{0pt}
\item \emph{Image canaries}: first $20$ distinct ScreenSpot v2
screenshots \citep{wu2024osatlas}; target
\texttt{CANARY-\{16hex\}-END} where the hex is derived from
\texttt{sha256("whack-canary-\{i:03d\}")}.
\item \emph{Text canaries}: per-canary user prompts ``Provide the
diagnostic label for sample id \{8hex\}'' paired with a shared
$224\!\times\!224$ mid-gray placeholder image; target
\texttt{TEXTCAN-\{16hex\}-END}.
\end{itemize}
Manifests are SHA-256-pinned on first run, verified each subsequent run.
Image-canary screenshots are excluded from bSFT data by filename
(perceptual-hash duplicate check in App.~\ref{app:overlap}).

\paragraph{Three injection regimes.}
On Qwen2.5-VL-7B-Instruct \citep{bai2025qwen25vl} with frozen vision
tower we inject canaries via LoRA \citep{hu2022lora} on the language
tower (target modules
$\{q,k,v,o,\text{gate},\text{up},\text{down}\}$\_proj), then
\texttt{merge\_and\_unload} the canary LoRA into the base \emph{before}
bSFT (``stacked LoRA bSFT'' $=$ canary-merged base + bSFT-LoRA loaded
at probe time; App.~\ref{app:mergeunload}). All canary LoRAs use
$r\!=\!32$, $\alpha\!=\!64$:
\begin{itemize}\setlength{\itemsep}{1pt}\setlength{\parskip}{0pt}
\item $M_{\mathrm{canary}}$ (image, saturated): $80$ ep, recall $20/20$.
\item $M_{\mathrm{text\,canary}}$ (text, saturated): $40$ ep, recall $20/20$.
\item $M_{\mathrm{canary\,under}}$ (image, undertrained): $10$ ep, baseline mean NLL $1.28$, hex $2.62$, hit $0/20$ (headroom for reactivation).
\end{itemize}

\paragraph{Non-canary-string SFT (bSFT) variants.}
``Benign'' here denotes the operational property that the SFT data
contains no copy of any canary string; \textsc{Safety} and
\textsc{Random} share the gray-placeholder image with text canaries
by construction, only the canary strings are excluded. On each canary-injected base we
stack a second LoRA ($r\!=\!16$, $\alpha\!=\!32$, lr $2\!\times\!10^{-5}$,
batch $2$, grad-accum $4$, cosine, $5000$ examples) on one of four data
sources (none contain canary strings):
\begin{itemize}\setlength{\itemsep}{1pt}\setlength{\parskip}{0pt}
\item \textsc{GUI}: ScreenSpot grounding (image$+$text).
\item \textsc{SYNTH}: PIL-generated abstract images with templated
captions (image$+$text out-of-domain control).
\item \textsc{Safety}: \texttt{(harmful-request, refusal)} text pairs
with placeholder image.
\item \textsc{Random}: placeholder-image factoid-question / short-answer
text pairs.
\end{itemize}
Matched-norm \textsc{LoRA-Noise} controls cover image, image-SYNTH,
T-GUI, T-Safety, U-image ($14$ aggregate cells).
Step grid (asymmetric, see Tabs.~\ref{tab:allcells} and~\ref{tab:accounting}):
\textsc{GUI}-saturated-image at $\{1,3,5,7,10\}$k; other saturated at
$\{1,3,5\}$k; undertrained-image at $\{1,3\}$k; $3$ seeds per cell.

All training and probes use bf16; App.~\ref{app:algebra} reports the bf16
residual between $K\!=\!20$ probe-pass and $K\!=\!30$ token-level pass for
completeness.

\paragraph{Reproducibility fields.}
Optimizer AdamW, peak lr $2\!\times\!10^{-5}$, cosine schedule, batch
$2$ with grad-accum $4$ (effective $8$), max sequence length $1024$,
canary LoRA epochs as listed above, bSFT seed-level training to $5000$
examples (or the cell-specific step grid). RNG seeds $\{0,1,2\}$ apply
jointly to bSFT data sampling, canary order, optimizer and dropout.
Decoding for hit$@k$ uses $T\!=\!0.7$, $p\!=\!0.95$, max new tokens
$=$ canary length $+\,4$, and the standard Qwen2.5/Llama-3 chat
template with the identical user prompt above; bSFT data is sampled
without replacement from each source. Cross-family runs use the same
text-only canary subset and the same hyperparameters, modulo
removing image fields from the canary template and the bSFT
batches.

\paragraph{Tokenization and probe windows.}
The text-canary assistant target tokenizes deterministically
(Table~\ref{tab:tokenization}); the template extends past token $20$.
Two consequences: (a) our internally pre-specified $K\!=\!20$ mean-NLL window covers
all $10$ preamble tokens, the literal, but only $9$ of $13$ hex tokens,
omitting the last $4$ hex tokens and the suffix; (b) the secret-span
(\texttt{canary\_hex}) probe widens the forward pass to $K\!=\!30$ to
include the full $13$ hex tokens.

\begin{table}[!h]
\centering
\small
\setlength{\tabcolsep}{4pt}
\begin{tabular}{@{}lccc@{}}
\toprule
\textbf{Span} & \textbf{Token positions} & \textbf{\# tokens} & \textbf{In $K\!=\!20$?} \\
\midrule
\texttt{preamble}      & $0$--$9$    & $10$ & all $10$ \\
\texttt{canary\_lit}   & $10$        & $1$  & yes \\
\texttt{canary\_hex}   & $11$--$23$  & $13$ & first $9$ only \\
\texttt{canary\_end}   & $24$        & $1$  & no (truncated) \\
\texttt{trailing}      & $25$+       & --   & no (truncated) \\
\bottomrule
\end{tabular}
\caption{Token-position layout of the \emph{text}-canary target
(Qwen2.5 tokenizer, zero-indexed). The image-canary target
\texttt{...labeled CANARY-\{16hex\}-END containing $N$ colored shapes...}
uses the same span definitions; \texttt{canary\_lit} is the
single token \texttt{ARY-} when present (in $\sim\!70\%$ of image
canaries the leading \texttt{CAN} is BPE-absorbed into preamble; in the
other $\sim\!30\%$ BPE merges \texttt{CAN} with \texttt{ARY-} so
\texttt{canary\_lit} is empty inside $K\!=\!20$, giving the $\sim\!0.70$
average count used in Appendix~\ref{app:algebra}). In both templates the
last hex tokens and the suffix fall outside $K\!=\!20$; per-canary
position layouts are deterministic given the SHA-derived hex string.}
\label{tab:tokenization}
\end{table}

\paragraph{Notation summary.}
$\mathrm{mean}_{K}$ always denotes a prefix-window average over
positions $0,\ldots,K\!-\!1$; $\mathrm{mean}_{20}$ is the internally
pre-specified prefix probe.  $\mathrm{hex}$ denotes the fixed
$13$-token secret span (positions $11$--$23$), evaluated with a
$K\!=\!30$ forward pass only to expose the tail tokens; it is not a
$30$-token average.  A $K\!=\!30$ pass is used only to obtain token
losses beyond position~$19$; unless explicitly written as
$\mathrm{mean}_{30}$, all $\mathrm{hex}$ values are averages over the
$13$ secret tokens only.  All $\Delta$ quantities are
bSFT-vs.-canary-baseline deltas in NLL loss units (additive, not
ratios), aggregated as the per-canary mean averaged across
$n\!=\!3$ outer seeds.

\paragraph{Three probes.}
Token positions zero-indexed throughout (Tab.~\ref{tab:tokenization});
$K\!=\!20$ $=$ positions $0$--$19$. Let
$\ell_\theta(t_j)\!=\!-\log p_\theta(t_j\!\mid\!t_{<j},x)$ and
$\ell^{\mathrm{base}}_\theta$ the per-token cross-entropy under bSFT and
canary-injected baseline.
$\Delta\mathrm{NLL}_{\mathrm{mean20}}\!=\!\frac{1}{20}\sum_{j=0}^{19}[\ell_\theta(t_j)-\ell^{\mathrm{base}}_\theta(t_j)]$;
the truncated-window mean-NLL probe and the $K\!=\!20$ cut are our own
design choice. $\Delta\mathrm{NLL}_{\mathrm{hex}}$ averages
the $13$ \texttt{canary\_hex} tokens (positions $11$--$23$; $11$--$19$
inside $K\!=\!20$) with a $K\!=\!30$ forward pass; $\Delta$ on other
spans defined analogously. Throughout the paper, $\mathrm{mean}_{K}$
denotes a prefix-window average over positions $0\!-\!(K\!-\!1)$,
while $\mathrm{hex}$ denotes the fixed $13$-token secret-span average,
even when evaluated with a $K\!=\!30$ forward pass for tail-token
coverage.
hit$@k$ is the fraction of canaries whose literal target appears in
$k\!\in\!\{1,4,16\}$ decoded outputs: greedy at $k\!=\!1$ and greedy
plus $k\!-\!1$ stochastic samples at $T\!=\!0.7,p\!=\!0.95$ for
$k\!\in\!\{4,16\}$ (i.e.\ $3$ and $15$ samples beyond greedy).

\paragraph{Aggregation and uncertainty.}
Each cell pools $20$ canaries $\times\,3$ seeds. Main-text statistics are
reported as the per-canary mean delta with a $95\%$ \emph{hierarchical
bootstrap} CI (outer-resample $3$ seeds with replacement, inner-resample
$20$ canaries within each chosen seed, $B\!=\!10{,}000$, fixed RNG seed).
The hierarchical bootstrap accounts for the fact that the same $20$
canaries are reused across seeds, unlike a naive IID bootstrap over
$60$ measurements. With only $3$ seeds these CIs are still wide; we
report seed ranges in addition wherever a claim is tight, and we do not
treat narrow CIs as population-level guarantees.
Exact bit-for-bit reproduction is not possible from this submission
alone because canary manifests and bSFT split files are not released;
the printed tables are intended for numeric audit, not rerunning, and
the appendix recipe in App.~\ref{app:repro} documents the construction
without releasing the artifacts.

\begin{table}[!t]
\scriptsize
\centering
\setlength{\tabcolsep}{3pt}
\begin{tabular}{@{}lrrr@{}}
\toprule
\textbf{Cond.} & \textbf{$\Delta\mathrm{NLL}_{\mathrm{mean20}}$} & \textbf{$\Delta\mathrm{NLL}_{\mathrm{hex}}$} & \textbf{hit$@1$} \\
\midrule
C1 img-\textsc{GUI} 10k & $+0.0015$ & $+0.0029$ & $1.00$ \\
C2 img-\textsc{SYNTH} 5k  & $+0.0004$ & $+0.0003$ & $1.00$ \\
\textbf{C3 txt-\textsc{GUI} 5k}  & $+0.0001$ & $\mathbf{+0.0133}$ & $\mathbf{0.88}$ \\
C4 txt-\textsc{Safety} 5k  & $+0.0150$ & $+0.0004$ & $1.00$ \\
C5 U-\textsc{GUI} 3k    & $-0.0070$ & $+0.0076$ & $0.00$ \\
\bottomrule
\end{tabular}
\caption{\textbf{Probe dissociation across five canonical cells ($\Delta$
point estimates).}
Qwen2.5-VL-7B, $20$ canaries $\times\,3$ seeds. Canary-injected
baselines (absolute NLL, not deltas): img$\!=\!0.0002/0.0002$,
txt$\!=\!0.0001/0.0003$, U-img$\!=\!1.2824/2.6184$
(mean$_{20}$/hex). Hierarchical-bootstrap and Student-$t_2$ $95\%$
intervals (descriptive at $n\!=\!3$) in App.~\ref{app:perseed} and
App.~\ref{app:allcells}. \textbf{C3} (bold):
unique saturated cell with hit$@1\!<\!1$ (Case~1; mean$_{20}$ misses it).
C4: drift $\sim\!99\%$ on \texttt{preamble} (Case~2). C5: in-$K{=}20$ hex
$-0.00583$ ($\sim\!83\%$) $+$ \texttt{canary\_lit} $-0.00091$ ($\sim\!13\%$)
$+$ bf16 residual $-0.00030$ $=$ canonical $-0.00703$
(Case~3; App.~\ref{app:algebra}). Full-span hex on C5 is positive in
point and across all $3$ seeds; descriptive intervals
($t_2$ excludes zero, hb brackets zero) reported in
App.~\ref{app:perseed}.}
\label{tab:dissociation}
\end{table}

\section{Three probe-disagreement case studies}
\label{sec:results-failures}

We surface three \emph{post-hoc} case studies (\S\ref{ss:case1}--\S\ref{ss:case3}, C3/C4/C5).
The matrix is $4$ bSFT variants $\times\,3$ canary regimes (saturated image,
text, undertrained image) $\times\,3$ seeds $\times\,3$--$5$ steps,
yielding $34$ aggregate bSFT cells $=$ $102$ seed-level runs
($144$ with $42$ \textsc{LoRA-Noise} controls).
Table~\ref{tab:dissociation} shows five canonical cells; full
all-cells point estimates in App.~\ref{app:allcells}; headline-cell
hb and $t_2$ intervals are listed in App.~\ref{app:perseed}.

\paragraph{Selection audit.}
Case studies are post-hoc (full audit:
Tab.~\ref{tab:selection}/App.~\ref{app:selection}). Headline counts:
\begin{itemize}\setlength{\itemsep}{1pt}\setlength{\parskip}{0pt}
\item \textbf{C3} sole hit$@1\!<\!1$ saturated cell ($26$-cell pool).
\item \textbf{C5} largest post-hoc negative mean$_{20}$ among $8$
undertrained-image bSFT cells ($3/3$ seeds same sign).
\item \textbf{C4} largest of $3$ qualifying preamble drift cells.
\item Across $48$ CI rows, $0$ cells have hb-CI screened negative on
$\Delta_{\mathrm{hex}}$.
\end{itemize}
Thresholds were chosen before per-cell CIs were aggregated, but
were \emph{not} preregistered or externally certified (App.~\ref{app:repro}).

\paragraph{Statistical note ($n\!=\!3$).}
Both hierarchical-bootstrap and Student-$t_2$ intervals are descriptive
only; per-seed sign-consistency describes the selected cells and is
\emph{not} inferential evidence after post-hoc selection
(App.~\ref{app:perseed}, App.~\ref{app:scope}).

\paragraph{``Benign'' caveat.}
``Benign'' here means operationally ``no canary string in bSFT data,''
not semantically benign. T-\textsc{Safety}/C4 share a fixed gray
placeholder image with text canaries (read as ``shared-placeholder'').

\subsection{Case 1 --- window truncation misses sparse greedy tail-token brittleness (C3).}
\label{ss:case1}

\textbf{Numbers.}
$\Delta_{\mathrm{mean20}}\!=\!+0.0001$ flat;
$\Delta_{\mathrm{hex}}\!=\!+0.0133$ ($3/3$ seeds positive); hit$@1$
drops $1.00\!\to\!0.88$.

\textbf{Mechanism.} Per-token plot (Fig.~\ref{fig:pertoken}): tokens
$0$--$22$ track baseline within $10^{-4}$ in absolute NLL; token $23$
(the final \texttt{canary\_hex} BPE piece, $1$--$2$ hex characters
before \texttt{-END}) absolute NLL jumps to $\sim\!0.119$ (per-token
mean across $60$ instances), three orders of magnitude above the
flat $\sim\!10^{-4}$ baseline at neighbouring positions. Position $23$
falls outside $K\!=\!20$ (zero-indexed), so mean$_{20}$ is blind to it.

\textbf{Sparsity.} $7/60$ greedy failures on $3/20$ canaries
(canary~$0$ dominates: $+0.122$ alone, leave-one-out hex $+0.0075$);
$17/20$ canaries perfect across all seeds; hit$@16\!=\!58/60$
(App.~\ref{app:c3failures}). One sparse tail-token counterexample, not
broad damage and not distribution-level loss of the secret.

\subsection{Case 2 --- non-secret preamble drift (C4).}
\label{ss:case2}

\textbf{Numbers.} $\Delta_{\mathrm{mean20}}\!=\!+0.0150$ ($3/3$ seeds
positive); $\Delta_{\mathrm{hex}}\!=\!+0.0004$; hit$@1\!=\!1.00$.

\textbf{Primary diagnostic (span decomposition).} Per-span decomposition
(App.~\ref{app:algebra}) attributes $\sim\!99\%$ of the mean$_{20}$
movement to non-secret \texttt{preamble} ($+0.01486/+0.01503$
token-level); only $+0.00017$ falls on \texttt{canary\_hex}. A probe
whose mass sits on non-secret preamble is not secret-span damage.

\textbf{Appendix-only sensitivity check (merge-and-unload).} Merging bSFT-LoRA
into the base collapses $\Delta_{\mathrm{preamble}}$ from $+0.0297$
to $+0.0014$ ($-95\%$); C1/C3/C5 secret-span ratios stay
$0.98/0.98/0.77$ and C4's own \texttt{canary\_hex} survives at $0.51$
(Tab.~\ref{tab:mergeunload}). The collapse is C4-preamble-specific
$\Rightarrow$ evaluation-stack-dependence on preamble (no causal claim).
This ablation is diagnostic of evaluation-stack sensitivity only; the
primary evidence for Case~2 is the per-span decomposition above.

\subsection{Case 3 --- unresolved undertrained-control ambiguity (C5).}
\label{ss:case3}

\textbf{Numbers.} $\Delta_{\mathrm{mean20}}\!=\!-0.0070$ ($3/3$ seeds
negative); full-span $\Delta_{\mathrm{hex}}\!=\!+0.0076$;
hit$@1\!=\!0$.

\textbf{Decomposition} (App.~\ref{app:algebra}): in-window hex
$-0.00583$ + \texttt{canary\_lit} $-0.00091$ + bf16 residual
$-0.00030$.

\textbf{Reading.} An in-window hex-token NLL drop on an undertrained
baseline; \emph{no} secret-specificity claim. Format-prior smoothing or
template adaptation are equally consistent without true-vs-decoy probes.

\paragraph{Per-seed sign consistency for headline cells.}
\begin{center}\scriptsize
\setlength{\tabcolsep}{4pt}
\begin{tabular}{@{}llrrrc@{}}
\toprule
\textbf{Cell} & \textbf{Probe} & \textbf{seed 0} & \textbf{seed 1} & \textbf{seed 2} & \textbf{Sign} \\
\midrule
C3 & $\Delta_{\mathrm{hex}}$    & $+0.0151$ & $+0.0146$ & $+0.0102$ & $3/3\!+$ \\
C4 & $\Delta_{\mathrm{mean20}}$ & $+0.0130$ & $+0.0161$ & $+0.0161$ & $3/3\!+$ \\
C5 & $\Delta_{\mathrm{mean20}}$ & $-0.0063$ & $-0.0082$ & $-0.0066$ & $3/3\!-$ \\
\bottomrule
\end{tabular}
\end{center}

\paragraph{C5 algebra closure.}
\[
\begin{aligned}
&\underbrace{-0.00583}_{\text{in-K hex}}
+ \underbrace{-0.00091}_{\text{canary\_lit}}\\
&\quad + \underbrace{+0.00001}_{\text{preamble}}
+ \underbrace{-0.00030}_{\text{bf16}}
= -0.00703 .
\end{aligned}
\]

\subsection{Window K-sweep: 7B failure modes vs.\ smaller-family stress test}
\label{ss:ksweep}

For the smaller-family stress test (\S\ref{ss:cross-family}) we
re-aggregate the per-token NLL files at $K\!\in\!\{10,15,20,25,30\}$;
for the 7B headline cells we report only the derivable extremes
$K\!=\!10$, $K\!=\!20$, and the full \texttt{canary\_hex} span (the
intermediate 7B $K\!\in\!\{15,25,30\}$ values would require re-running
the per-token aggregator with a different window length and are not
printed in this submission).

\textbf{On the 7B headline cells}, the algebra closure
(Tab.~\ref{tab:algebra}) plus the full-span readout
(Tab.~\ref{tab:dissociation}) pin down the printed extremes (see
Tab.~\ref{tab:ksweep7b}). They imply: \textbf{C3} is
$\sim\!+10^{-4}$ for $K\!\leq\!20$ and only diverges from the in-window
value once the tail-token spike at position~$23$ enters the window;
\textbf{C4} stays positive across $K\!=\!10\!\to\!20$, with the
preamble-only $K\!=\!10$ mean ($+0.0297$) accounting for
$\sim\!99\%$ of the canonical $K\!=\!20$ value after rescaling by
$10/20$ (i.e.\ $0.5\!\times\!0.0297\!=\!0.01486$ vs.\ $+0.01504$);
\textbf{C5}'s mean is negative at the canonical $K\!=\!20$ but turns
positive at the full-span readout, the joint-product signature of
window truncation \emph{and} the undertrained-baseline regime.

\begin{figure}[!t]
\centering
\IfFileExists{figures/fig_ksweep_600dpi_cropped.png}{%
  \includegraphics[width=\linewidth]{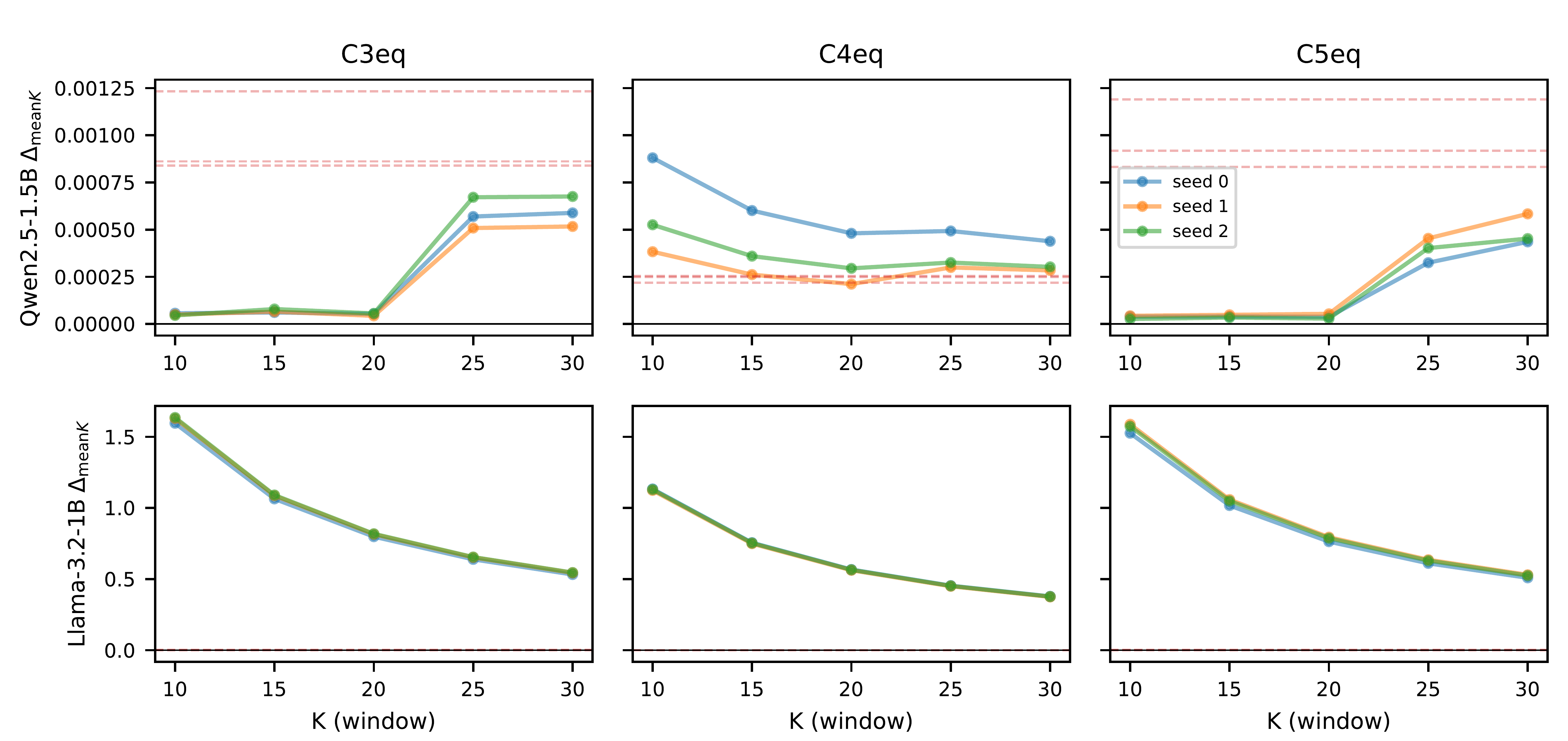}%
}{%
\IfFileExists{figures/fig_ksweep_600dpi.png}{%
  \includegraphics[width=\linewidth,trim=7pt 5pt 7pt 1pt,clip]{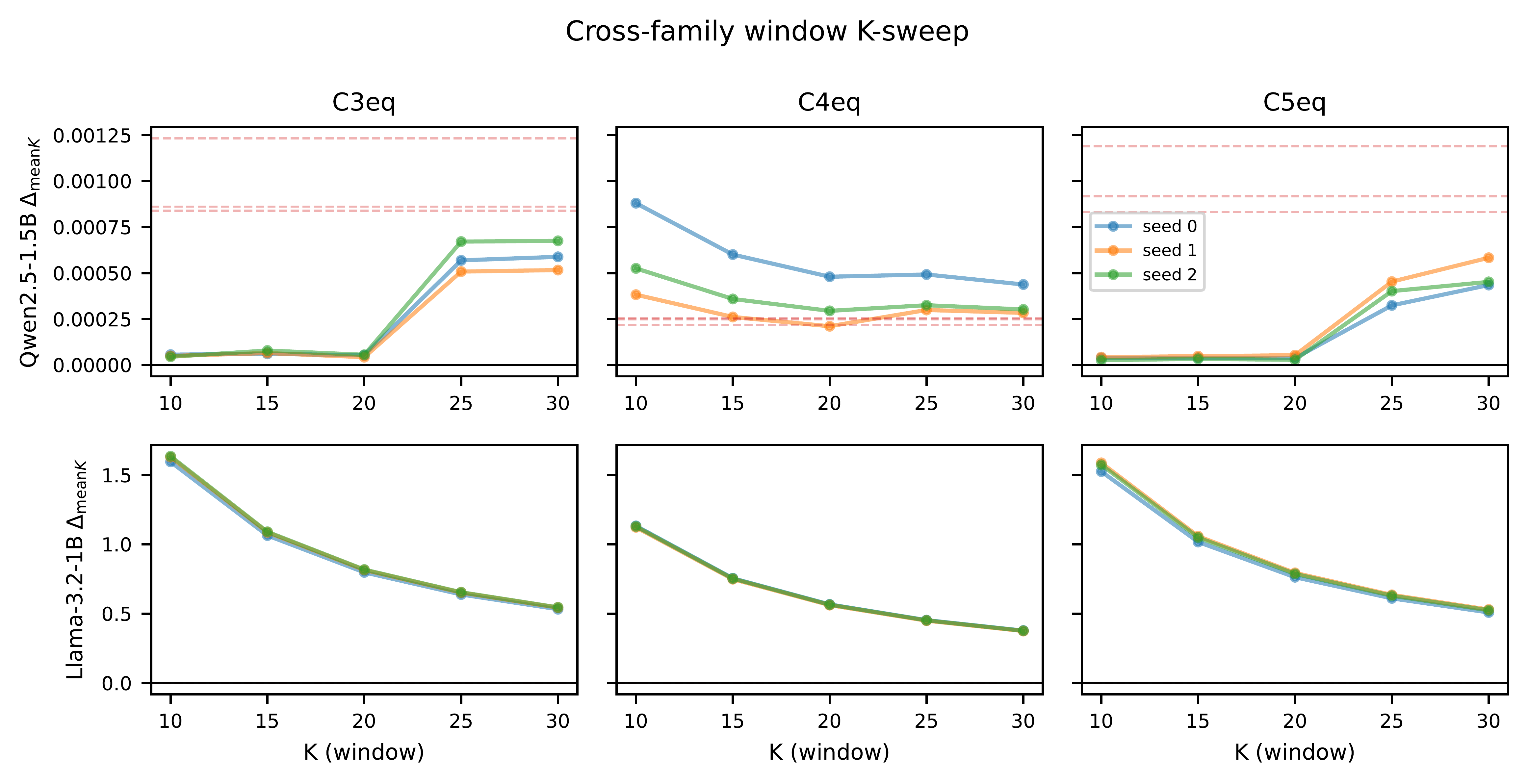}%
}{%
  \fbox{\parbox{0.9\linewidth}{\centering\itshape
  Placeholder: figures/fig\_ksweep.pdf to be produced by
  \texttt{scripts\_fogen/run\_phase0\_ksweep.sh}.}}%
}%
}
\caption{\textbf{Descriptive smaller-family stress test.}
$\Delta\mathrm{NLL}_{\mathrm{mean}K}$ for $K\!\in\!\{10,15,20,25,30\}$
on Qwen2.5-1.5B (top) and Llama-3.2-1B (bottom) at the
C3/C4/C5-equivalent cells, per seed. Dashed red lines mark full
\texttt{canary\_hex} span $\Delta$ where present. Y-axis scales differ
by family by $\sim\!1000\times$; read this as a smaller-family stress
test, not as case-level replication.}
\label{fig:ksweep}
\end{figure}

\textbf{On the smaller-family C3/C4/C5-equivalent cells} (Fig.~\ref{fig:ksweep}):
the K-monotone shape on Qwen2.5-1.5B and Llama-3.2-1B differs from the
7B testbed: in particular, the negative-mean signature behind 7B
Case~3 does \emph{not} reproduce on either smaller family at the
C5-equivalent cell, and Llama-1B's $\Delta\mathrm{mean}_{K}$ at
C3/C4-eq is dominated by a catastrophic preamble drift at every $K$
rather than by the tail-token / preamble-fraction mechanism we observe
on 7B.

\subsection{Descriptive smaller-family non-replication check}
\label{ss:cross-family}

We run a smaller-family stress test at the C3/C4/C5-equivalent cells
on Qwen2.5-1.5B-Instruct~\citep{qwen25technical} and
Llama-3.2-1B-Instruct~\citep{meta2024llama32} (text-only LMs,
same canary template, same bSFT data minus image fields, same probe
family, $n\!=\!3$ outer seeds at each cell). We do not print or claim
a full $4\!\times\!3$ smaller-family bSFT grid; App.~\ref{app:allcells}
is the 7B all-cells matrix only. Headline-cell point estimates appear
in Tab.~\ref{tab:cross-family}; we read these as descriptive stress
tests on smaller text-only backbones, not as case-level replication of
the 7B mechanisms (Tab.~\ref{tab:cross-family} caption).

\begin{table}[t]
\centering\small
\setlength{\tabcolsep}{3pt}
\caption{\textbf{Smaller-family stress test on the C3/C4/C5-equivalent cells.}
Point-estimate $\Delta\mathrm{NLL}$ (mean$_{20}$ and full
canary\_hex span) and greedy hit$@1$ at $n\!=\!3$ outer seeds.
The 7B row reproduces Tab.~\ref{tab:dissociation}; the 1.5B / 1B rows
are descriptive stress-test runs on smaller text-only LMs at cells
fixed by the 7B labelling, not case-level replication of the 7B
mechanisms.  Qwen-1.5B C3/C4/C5-eq are near-zero with greedy
hit$@1\!=\!1.00$ throughout, and Llama-1B exhibits a different
catastrophic-mean / preserved-hex / broken-greedy disagreement pattern
at every cell.}
\label{tab:cross-family}
\begin{tabular}{l l rrr}
\toprule
\textbf{Family} & \textbf{Cell} & \textbf{$\Delta_{\mathrm{mean20}}$} & \textbf{$\Delta_{\mathrm{hex}}$} & \textbf{hit$@1$} \\
\midrule
Qwen2.5-VL-7B & C3-eq & $+0.0001$ & $+0.0133$ & $0.88$ \\
Qwen2.5-1.5B & C3-eq & $+0.0001$ & $+0.0010$ & $1.00$ \\
Llama-3.2-1B & C3-eq & $+0.8104$ & $+0.0006$ & $0.37$ \\
\midrule
Qwen2.5-VL-7B & C4-eq & $+0.0150$ & $+0.0004$ & $1.00$ \\
Qwen2.5-1.5B & C4-eq & $+0.0003$ & $+0.0002$ & $1.00$ \\
Llama-3.2-1B & C4-eq & $+0.5643$ & $+0.0001$ & $0.00$ \\
\midrule
Qwen2.5-VL-7B & C5-eq & $-0.0070$ & $+0.0076$ & $0.00$ \\
Qwen2.5-1.5B & C5-eq & $+0.0000$ & $+0.0010$ & $1.00$ \\
Llama-3.2-1B & C5-eq & $+0.7814$ & $+0.0006$ & $0.00$ \\
\bottomrule
\end{tabular}
\end{table}

\paragraph{Cross-family verdict.}
The smaller-family stress test surfaces \emph{additional} probe
disagreements but does \emph{not} replicate all three 7B case
mechanisms at the C3/C4/C5-equivalent cells.  Case~3's
undertrained-baseline mechanism has no direct analog in our text-only
cross-family setup (text LMs in this protocol have no separate
undertrained-canary regime; the image-canary gray-placeholder
construction that triggers C5 on the VLM testbed is unavailable on
text-only LMs), and the C5 cell at the 7B-fixed location does not
reproduce as a misleading negative on either family
(Tab.~\ref{tab:cross-family}: Qwen-1.5B C5-eq
$\Delta_{\mathrm{mean20}}\!=\!{+}0$; Llama-1B C5-eq $+0.78$ matches
the same catastrophic-mean / preserved-hex / broken-greedy pattern
as Llama C3-eq and C4-eq, not Case~3).  Two additional smaller-family
phenomena --- a migrated Qwen-1.5B C4-equivalent cell with $98\%$
preamble-driven $\Delta_{\mathrm{mean20}}\!=\!{+}0.22$, and a
Llama-3.2-1B T-bSFT-GUI step-trajectory where $\mathrm{mean}_{K=20}$
grows monotonically while greedy hit$@16$ recovers across step --- and
the smaller-family decoy audit at the candidate C5-equivalent cells
(showing additive NLL gaps of $+1.67$ to $+10.86$ loss units between
shuffled-hex / wrong-secret / format-only decoys and the true
target) are reported in App.~\ref{app:crossfamily-decoy}; this is a
calibration of the decoy construction at \emph{different} cells and
does not adjudicate the original 7B C5 ambiguity, which lacks a
paired decoy audit.  Magnitudes are
3-seed point estimates, not inferential, and we read these rows as
descriptive smaller-family stress tests rather than as case-level
replication.

\section{Related Work}
\label{sec:related}

\paragraph{Memorization probes for autoregressive models.}
The canary-and-probe paradigm we use originates with Secret Sharer
\citep{carlini2019secret}, which introduced \emph{exposure} as a
calibrated rank-based metric for unintended memorization, and was
generalized in \citet{carlini2023quantifying} to discoverable-extraction
rates across model scales, data duplication, and prompt length.
Truncated mean-NLL over the first $K$ target tokens --- the probe we
audit --- is a natural prefix-window operationalization in this
lineage; the present paper's three case studies show that, on a
single-backbone testbed, the choice of probe (truncated mean NLL,
full-span loss, behavioural exact-recall) can materially change
false-positive and false-negative rates. We do not propose a new probe; we document, in one
autoregressive generative-model testbed, three concrete cases where
truncated mean-NLL silently disagrees with full-span NLL or with
behavioral exact-recall, and attribute each disagreement to a
specific token-position or baseline-regime mechanism.

\paragraph{Extraction and large-scale memorization.}
\citet{carlini2021extracting} demonstrated that training data can be
recovered from large language models via query-based extraction with
ranking-based filtering, shifting the field toward extraction-based memorization metrics that
do not depend on a fixed-window NLL probe at all; subsequent work
\citep{nasr2023scalable} scales these attacks to production language
models. Memory-trace studies that frame memorization as a
finetuning-time and forgetting-time quantity
\citep{jagielski2023measuring} sit between extraction-based and
NLL-based metrics and motivate the need to know what fixed-window NLL
does and does not measure on a controlled canary set.
Deduplication-driven studies
\citep{lee2022deduplicating,carlini2023quantifying} show that duplicate
training text affects memorized emissions and that memorization
increases with duplication count, linking training distribution to probe
response. Our setting is adjacent: the canary distribution is
deliberately controlled, and we audit the probe rather than the
underlying memorization rate, but the same evaluation-metric concerns
apply.

\paragraph{Broader evaluation-reliability audits.}
Our probe-disagreement view is part of a broader reliability concern:
diagnostics, explanations, domain-validity audits, and benchmark verdicts
can shift when the measured object changes
\citep{lan-etal-2025-attention,zhuang2026preregistering,li2026safetyrepro,wang2026auditing}.
Adjacent generative-model and RAG evaluations make a similar point through
multidimensional bias benchmarks, context-conflict probes, and
evidence-warrant audits
\citep{luo2026biasig,luo2024faintbench,chen2026doesragknowretrieval,qian2026relevantwarrantedevidenceforcecalibration}.
We cite them as measurement-reliability neighbors rather than
memorization-probe baselines.

\paragraph{Benign or safety-driven side effects on memorization.}
The closest motivating works are \citet{liu2026whack}, on
Whack-a-Mole-style recall activation after benign or task-style
fine-tuning, and \citet{borkar2025ripple}, on privacy-ripple effects
from adding or removing personal information.
These claims are exactly the kind of memorization-versus-generalization
question that probe choice can flip --- the same probe-window decision
that buries Case~1 (a tail-token greedy miss outside $K\!=\!20$) can
also bury any reactivation that occurs predominantly outside the
chosen window. We do not claim our three case studies refute or
replicate those reactivation findings; we use a related probe setting
on a different testbed and document where it is and is not
self-consistent.

\paragraph{Unlearning and lifecycle.}
A separate line studies training-data unlearning, from general
SISA-style removal \citep{bourtoule2021unlearning} to targeted
forgetting of memorized generative-model content \citep{maini2024tofu}.
Recent LLM unlearning audits and unlearning-adjacent recommender work
sharpen the same measurement question: forgetting should be tied to the
measured object \citep{li2026auditing}, while recommender unlearning
mechanisms provide an adjacent measured-object-specific design
\citep{chen2026cure}. The probe-disagreement phenomena we document are a
precondition for unlearning evaluation: if truncated mean-NLL drops on an
undertrained baseline without secret-specific evidence (Case~3), audits
using the same probe family may register apparent ``forgetting'' without
a behavioural change. We adapt the decoy/canary control of Secret Sharer
\citep{carlini2019secret} to expose this ambiguity.

\paragraph{Memorization in non-autoregressive generative models.}
Memorization in diffusion models is an active area of the foundations
literature
\citep{carlini2023diffusion,somepalli2023diffusion,somepalli2023understanding},
and the metric-dependence concern we document is not specific to
autoregressive backbones: visual generative models can be probed via
near-duplicate retrieval, likelihood-based membership inference, or
direct extraction, each of which has its own false-positive and
false-negative profile relative to true training-image replication. We do not run experiments on
diffusion models in this paper; we limit our claims to
an autoregressive setting and discuss in \S\ref{sec:discussion} which
parts of the failure-mode taxonomy plausibly carry over and which
require dedicated experiments.

\paragraph{What this paper adds.}
Relative to the works above, our contribution is narrow and
audit-shaped: in one testbed, we treat truncated mean-NLL as the unit of
analysis, decompose it onto explicit token spans
(\texttt{preamble}, \texttt{canary\_lit}, \texttt{canary\_hex},
\texttt{canary\_end}, \texttt{trailing}; Tab.~\ref{tab:tokenization}),
and report three cases where span or behavioural-recall verdicts diverge
from the truncated probe. Three cases in one backbone are not a
population claim; they make the failure modes \emph{nameable} so future
probe-window choices can be reported with the relevant audits attached.

\balance
\section{Discussion: evaluation protocols for memorization in generative models}
\label{sec:discussion}

\paragraph{Take-home.}
In this testbed, truncated $\mathrm{mean}_{K=20}$ NLL yields a false
negative (Case~1), a false positive (Case~2), and an ambiguity
(Case~3: an undertrained in-window NLL drop without full-secret
recovery, not a secret-specific recovery claim).
Memorization audits on canary sets should therefore report, beyond
$\Delta\mathrm{NLL}_{\mathrm{mean20}}$ alone:
\vspace{-8pt}
\begin{itemize}\setlength{\itemsep}{0pt}\setlength{\parsep}{0pt}\setlength{\parskip}{0pt}\setlength{\topsep}{0pt}\setlength{\partopsep}{0pt}
\item[\textbf{(a)}] \textbf{Full-span secret NLL.} A window that omits
part of the secret span is blind to omitted-position damage. Case~1:
the last hex BPE piece is at position~$23$, outside $K\!=\!20$, so a
$0\!-\!19$ window drops it.
\item[\textbf{(b)}] \textbf{Behavioural hit$@k$ with $k\!\geq\!4$.}
Greedy hit$@1$ catches Case~1's tail-token brittleness; sampled
hit$@k$ checks whether the secret persists. In our matrix,
hit$@16\!=\!58/60$ on C3, so broken canaries remain in the secret
distribution.
\item[\textbf{(c)}] \textbf{Per-span decomposition (e.g.\
\texttt{preamble} vs.\ secret).} Isolates whether a probe move is on
the secret or on template-shared positions. Case~2's $+0.0150$ probe
move sits $\sim\!99\%$ on the non-secret \texttt{preamble} span while
the secret-span $\Delta$ is $+0.0004$ and hit$@1\!=\!1.00$.
\item[\textbf{(d)}] \textbf{A saturated and an undertrained control}
to bound how far reactivation could move the probe in principle.
Across the $4\!\times\!2$ undertrained-image bSFT grid plus
matched-norm \textsc{LoRA-Noise}, no variant produces verbatim
hit$@1\!>\!0$ (App.~\ref{app:scope}), so probe drops in this regime
cannot be attributed to secret recovery without further evidence.
\item[\textbf{(e)}] \textbf{Decoy probes (shuffled / wrong-secret /
format-only)} before asserting secret-specificity. Case~3: an
in-window hex-token NLL drop is consistent with both secret-token
learning and format-prior smoothing. A 7B C5 decoy remains future
work (cf.\ App.~\ref{app:crossfamily-decoy}).
\end{itemize}



\clearpage
\bibliographystyle{icml2026}
\begingroup
\sloppy
\emergencystretch=2em
\bibliography{references}
\endgroup

\onecolumn
\appendix
\raggedbottom
\section{Scope of claims and reproducibility}
\label{app:scope}

\paragraph{Scope of claims.}
The three case studies hold in our testbed only: Qwen2.5-VL-7B-Instruct,
language-tower LoRA with frozen vision tower, four bSFT variants
(\textsc{GUI}, \textsc{SYNTH}, \textsc{Safety}, \textsc{Random}), $20$
canaries per modality (image, text, undertrained-image), exact substring
probes, $K\!=\!20$ for mean-NLL. Across the $4\!\times\!2$
undertrained-image bSFT grid plus matched-norm \textsc{LoRA-Noise}
($2$ steps), no variant produces verbatim hit$@1\!>\!0$ on the
undertrained canaries. The strongest evidence (Case~1, C3) is
essentially text-only; we do not extrapolate to other backbones,
PEFT configurations, canary templates, or non-autoregressive families
(diffusion, flow). $n\!=\!3$ seeds; hierarchical-bootstrap and
Student-$t_2$ intervals are descriptive only, and after post-hoc
selection no inferential claim survives selection adjustment
(Tab.~\ref{tab:selection}).

\paragraph{Reproducibility.}
\label{app:repro}
We do not release supplementary material. All canonical numbers are
reproduced in the printed tables: per-cell deltas in
Table~\ref{tab:allcells}, per-seed sign-consistency and headline
intervals in App.~\ref{app:perseed}, span decomposition in
Table~\ref{tab:algebra}, and merge/stacked comparisons in
Tables~\ref{tab:logit_agreement}--\ref{tab:mergeunload}.
Selection-audit thresholds in Table~\ref{tab:selection}
($|\Delta_{\mathrm{mean20}}|\!>\!10^{-3}$,
$|\Delta_{\mathrm{hex}}|\!<\!0.5|\Delta_{\mathrm{mean20}}|$, baseline $+$
post hit$@1\!=\!1$, screened negative under hb interval) were chosen
before per-cell CIs were aggregated but were \emph{not} preregistered
or externally certified; read them as \emph{authors' internal selection
thresholds}, not a registered analysis plan. The full pipeline is
recoverable from the body and tables; exact bit-for-bit reproduction
requires the unreleased canary manifests and bSFT splits. Canary
secrets are SHA-256 of \texttt{whack-canary-\{i:03d\}} truncated to
$16$~hex; bSFT samples $5000$ examples without replacement under seed
$s\!\in\!\{0,1,2\}$ with image-filename exclusion
(\S\ref{app:overlap}); training is stacked LoRA ($r\!=\!16$,
$\alpha\!=\!32$) on top of the canary-merged base with AdamW, peak lr
$2\!\times\!10^{-5}$, cosine schedule, batch
$2\!\times\!\mathrm{accum}\,4$, max sequence length $1024$.

\section{Selection audit (full)}
\label{app:selection}

\begin{table}[H]
\centering
\scriptsize
\setlength{\tabcolsep}{2.5pt}
\begin{tabular}{@{}lp{2.05cm}cp{2.4cm}@{}}
\toprule
\textbf{Family} & \textbf{Pool} & \textbf{\#qual.} & \textbf{Shown} \\
\midrule
hit$@1\!<\!1$ at saturated baseline & $26$ saturated cells (img$+$txt) & $1$ & C3 (unique) \\
\addlinespace[1pt]
preamble drift: $|\Delta_{\mathrm{mean20}}|\!>\!10^{-3}$, $|\Delta_{\mathrm{hex}}|\!<\!0.5|\Delta_{\mathrm{mean20}}|$, baseline $+$ post hit$@1\!=\!1$ & $26$ saturated cells & $3$ (T-Safety $1\!/\!3\!/\!5$k) & C4 (largest magnitude) \\
\addlinespace[1pt]
undertrained-regime negative mean$_{20}$: $\Delta_{\mathrm{mean20}}\!<\!-10^{-3}$ & $8$ undertrained-image bSFT cells & $2$ (U-GUI 1k$/$3k) & C5 (largest magnitude $\!=\!$U-GUI 3k) \\
\addlinespace[1pt]
screened negative under hb interval on $\Delta_{\mathrm{hex}}$ & $48$-row CI dump & $0$ & --- \\
\addlinespace[1pt]
descriptively \emph{negative} hb interval on $\Delta_{\mathrm{mean20}}$ & $48$ & $1$ (U-GUI 3k$\!=\!$C5) & C5 (sole negative) \\
\bottomrule
\end{tabular}
\caption{\textbf{Selection audit.} Eligible pool, threshold, and qualifier
count for every diagnostic family we examined. Thresholds were
chosen before per-cell CIs were aggregated, not externally certified
(App.~\ref{app:repro}). \textbf{C3} is the only
hit$@1\!<\!1$ saturated cell; \textbf{C5} is selected as the largest post-hoc negative
mean$_{20}$ point estimate among the $8$ undertrained-image bSFT
cells; its hb interval is the only negative mean$_{20}$ interval
among the $48$-row CI pool, reported descriptively only; \textbf{C4} is the
largest of $3$ qualifying preamble-drift cells. \textbf{C1, C2} are
diagnostic baselines (largest-step saturated \textsc{GUI}; saturated
\textsc{SYNTH 5k}, sub-threshold). The full matrix is reported in
Table~\ref{tab:allcells}; we do not claim outcome-blind selection.
$\rho(\Delta_{\mathrm{mean20}},\Delta_{\mathrm{hex}})\!=\!0.82$ across
$34$ bSFT cells.}
\label{tab:selection}
\end{table}

\section{Per-seed sign-consistency and headline intervals}
\label{app:perseed}

Across the five canonical cells, every per-seed
$\Delta\mathrm{NLL}_{\mathrm{mean20}}$, $\Delta\mathrm{NLL}_{\mathrm{hex}}$,
and hit$@1$ value is sign-consistent across the $3$ outer seeds (max
spread $0.0049$ on C3 hex, $0.0255$ on C5 \texttt{canary\_lit}); the
3-seed mean of each cell equals Table~\ref{tab:dissociation} by
construction. Headline-row Student-$t_2$ vs.\ hier-bootstrap
intervals: \textbf{C3 hex} $[+0.0066,+0.0200]$ / $[+0.0061,+0.0226]$;
\textbf{C4 mean$_{20}$} $[+0.0106,+0.0195]$ / $[+0.0129,+0.0169]$;
\textbf{C5 mean$_{20}$} $[-0.0096,-0.0045]$ / $[-0.0118,-0.0021]$;
\textbf{C5 full hex} $[+0.0027,+0.0124]$ / $[-0.0045,+0.0198]$. Both
interval families are descriptive at $n\!=\!3$; we read claims as
joint requirements (same direction in both) and mark \emph{hb-only}
any claim that relies on hier-bootstrap alone.

\section{\texorpdfstring{Algebra closure: span decomposition of $\Delta\mathrm{NLL}_{\mathrm{mean20}}$}{Algebra closure: span decomposition of Delta NLL mean20}}
\label{app:algebra}

For each canonical cell we decompose $\Delta\mathrm{NLL}_{\mathrm{mean20}}$ into
per-span contributions \emph{inside} the $K\!=\!20$ window. For each (canary, seed)
instance let $n_{s,c}$ be the number of tokens of span $s$ that fall inside the
canary's $K\!=\!20$ window and $\overline{\Delta_{s,c}}$ the per-token
$\Delta\mathrm{NLL}$ over those tokens; the per-instance contribution of span $s$ is
$n_{s,c}\,\overline{\Delta_{s,c}}/20$ ($0$ if the canary has no tokens of $s$ inside
$K\!=\!20$). Averaging across all $60$ (canary, seed) instances and summing across
spans yields the per-canary mean $\Delta\mathrm{NLL}_{\mathrm{mean20}}$ from the
\emph{token-level} forward pass.

The canonical mean$_{20}$ uses the $K\!=\!20$ probe pass; the
token-level pass uses $K\!=\!30$ to cover the hex span. Both compute
teacher-forced cross-entropy in bf16 from the same weights, but bf16's
$7$-bit mantissa produces a per-cell residual of order $10^{-4}$
(max $-0.00030$ on C5). The residual is computed independently as
canonical minus enumerated and shown as a disclosure row, not a
free parameter. For \emph{text} canaries (C3, C4) the Qwen2.5
tokenizer yields exactly $n_{\mathrm{preamble}}\!=\!10$,
$n_{\mathrm{lit}}\!=\!1$, $n_{\mathrm{hex}}\!=\!9$ inside $K\!=\!20$;
for \emph{image} canaries (C1, C2, C5) BPE boundaries vary across
strings, with cell-level averages
$\overline{n}_{\mathrm{lit}}\!\approx\!0.70$,
$\overline{n}_{\mathrm{hex}}\!\approx\!9.30$.

\begin{table}[H]
\centering
\scriptsize
\setlength{\tabcolsep}{3pt}
\begin{tabular}{@{}llrrr@{}}
\toprule
\textbf{Cell} & \textbf{Span} & \textbf{$\overline{\Delta_s}$ /tok} & \textbf{$\overline{n_s}$} & \textbf{Contrib.} \\
\midrule
C1 (img) & \texttt{canary\_hex} & $+0.00317$ & $9.30$ & $+0.00148$ \\
 & \texttt{canary\_lit} & $-0.00004$ & $0.70$ & $-0.00000$ \\
 & \texttt{preamble} & $+0.00012$ & $10.00$ & $+0.00006$ \\
 & \emph{enumerated sum (token-level $K\!=\!30$)} & & $20.00$ & $+0.00154$ \\
 & \emph{bf16 precision residual} & & --- & $<\!10^{-5}$ \\
 & \textbf{canonical (Tab.~\ref{tab:dissociation})} & & $20.00$ & $\mathbf{+0.00154}$ \\
\addlinespace[2pt]
C2 (img) & \texttt{canary\_hex} & $+0.00041$ & $9.30$ & $+0.00019$ \\
 & \texttt{canary\_lit} & $+0.00004$ & $0.70$ & $+0.00000$ \\
 & \texttt{preamble} & $+0.00044$ & $10.00$ & $+0.00022$ \\
 & \emph{enumerated sum (token-level $K\!=\!30$)} & & $20.00$ & $+0.00041$ \\
 & \emph{bf16 precision residual} & & --- & $<\!10^{-5}$ \\
 & \textbf{canonical (Tab.~\ref{tab:dissociation})} & & $20.00$ & $\mathbf{+0.00041}$ \\
\addlinespace[2pt]
C3 (text) & \texttt{canary\_hex} & $+0.00013$ & $9.00$ & $+0.00006$ \\
 & \texttt{canary\_lit} & $+0.00000$ & $1.00$ & $+0.00000$ \\
 & \texttt{preamble} & $+0.00006$ & $10.00$ & $+0.00003$ \\
 & \emph{enumerated sum (token-level $K\!=\!30$)} & & $20.00$ & $+0.00009$ \\
 & \emph{bf16 precision residual} & & --- & $<\!10^{-5}$ \\
 & \textbf{canonical (Tab.~\ref{tab:dissociation})} & & $20.00$ & $\mathbf{+0.00009}$ \\
\addlinespace[2pt]
C4 (text) & \texttt{canary\_hex} & $+0.00037$ & $9.00$ & $+0.00017$ \\
 & \texttt{canary\_lit} & $+0.00000$ & $1.00$ & $+0.00000$ \\
 & \texttt{preamble} & $+0.02973$ & $10.00$ & $+0.01486$ \\
 & \emph{enumerated sum (token-level $K\!=\!30$)} & & $20.00$ & $+0.01503$ \\
 & \emph{bf16 precision residual} & & --- & $+0.00001$ \\
 & \textbf{canonical (Tab.~\ref{tab:dissociation})} & & $20.00$ & $\mathbf{+0.01504}$ \\
\addlinespace[2pt]
C5 (U-img) & \texttt{canary\_hex} & $-0.01253$ & $9.30$ & $-0.00583$ \\
 & \texttt{canary\_lit} & $-0.02601$ & $0.70$ & $-0.00091$ \\
 & \texttt{preamble} & $+0.00002$ & $10.00$ & $+0.00001$ \\
 & \emph{enumerated sum (token-level $K\!=\!30$)} & & $20.00$ & $-0.00673$ \\
 & \emph{bf16 precision residual} & & --- & $-0.00030$ \\
 & \textbf{canonical (Tab.~\ref{tab:dissociation})} & & $20.00$ & $\mathbf{-0.00703}$ \\
\bottomrule
\end{tabular}
\caption{Exact span decomposition of $\Delta\mathrm{NLL}_{\mathrm{mean20}}$ across the five canonical cells.
``(text)'' / ``(img)'' / ``(U-img)'' marks the canary modality
(text canaries deterministically tokenize to $n_{\mathrm{lit}}\!=\!1$,
$n_{\mathrm{hex}}\!=\!9$ inside $K\!=\!20$; image canaries are
sample-averaged at $\overline{n}\!\approx\!0.70/9.30$).
\textbf{Contrib.} is the per-instance contribution averaged across all $60$ (canary, seed)
instances ($0$ for canaries lacking that span inside $K\!=\!20$):
$\langle n_{s,c}\,\overline{\Delta_{s,c}}/20\rangle$. \textbf{Enumerated sum} reproduces the
per-canary mean of the token-level forward pass. \textbf{bf16 precision residual}
is computed independently as canonical minus enumerated; we display it as a
disclosure row, not as a free parameter (max $|\!\cdot\!|\!=\!0.00030$ on C5).
\textbf{Canonical} is identically the Table~\ref{tab:dissociation} value.
For \textbf{C5}, the $-0.0070$ canonical mean20 decomposes into a $\sim\!9.3$-token in-K=20
hex drop ($\sim\!-0.013$/tok, contribution $-0.00583$, $\sim\!83\%$ of canonical) and
a partial \texttt{canary\_lit} drop ($\sim\!-0.026$/tok where present in $\sim\!70\%$ of canaries,
contribution $-0.00091$, $\sim\!13\%$). The full-span hex point estimate
$\Delta\mathrm{NLL}_{\mathrm{hex}}\!=\!+0.0076$ in
Table~\ref{tab:dissociation} stays positive because out-of-K=20 hex tail
tokens degrade by $\sim\!+0.06$/tok on average; hit$@1\!=\!0.00$ across all
seeds. Sign-consistency: all $3$ seeds give negative mean$_{20}$ and
positive full-span hex.}
\label{tab:algebra}
\end{table}

\paragraph{7B window K-sweep --- derivable extremes.}
The K-monotone claims for the 7B headline cells are recoverable from
the algebra closure plus the headline full-span values: $K\!=\!10$ is
the preamble-only mean, $K\!=\!20$ is the algebra-closure mean, and
the full hex column reproduces Tab.~\ref{tab:dissociation}.
Intermediate $K\!\in\!\{15,25,30\}$ values require re-running the
per-token aggregator and are not printed.

\begin{table}[H]
\centering
\small
\setlength{\tabcolsep}{4pt}
\begin{tabular}{@{}lrrr@{}}
\toprule
\textbf{Cell} & $\mathrm{mean}_{K=10}$ (preamble-only) & $\mathrm{mean}_{K=20}$ & $\mathrm{hex}$ (full $13$-tok) \\
\midrule
C3 (txt-\textsc{GUI} 5k)  & $+0.0001$ & $+0.0001$ & $+0.0133$ \\
C4 (txt-\textsc{Safety} 5k)  & $+0.0297$ & $+0.0150$ & $+0.0004$ \\
C5 (U-\textsc{GUI} 3k)    & $+0.0000$ & $-0.0070$ & $+0.0076$ \\
\bottomrule
\end{tabular}
\caption{\textbf{7B window extremes for the headline cells (derivable from
the algebra closure).} $\mathrm{mean}_{K=10}$ is the per-canary
preamble-only mean ($10$ preamble tokens divided by $K\!=\!10$), so it
equals the per-token preamble $\overline{\Delta}_{\mathrm{preamble}}$
in Tab.~\ref{tab:algebra}; $\mathrm{mean}_{K=20}$ matches
Tab.~\ref{tab:dissociation}; the full \texttt{canary\_hex} column is
the per-canary mean over the $13$-token secret span (positions
$11$--$23$, $K\!=\!30$ pass). \textbf{C3} stays at $+10^{-4}$ for
$K\!\leq\!20$ and only diverges from the in-window value once the
tail-token spike at position~$23$ enters the window; \textbf{C4} is
positive at $K\!=\!10$ with the preamble already absorbing the bulk
and decays toward the small full-hex value as more hex positions
enter; \textbf{C5} is negative at $K\!=\!20$ but turns positive at the
full-hex readout. Point estimates only ($n\!=\!3$ outer seeds,
descriptive).}
\label{tab:ksweep7b}
\end{table}

\section{All-cells systematic table}
\label{app:allcells}

\begin{table}[H]
\centering
\scriptsize
\setlength{\tabcolsep}{3pt}
\begin{tabular}{@{}llrrrrr@{}}
\toprule
\textbf{Mod.} & \textbf{Variant} & \textbf{Step} & \textbf{$\Delta_{\mathrm{mean20}}$} & \textbf{$\Delta_{\mathrm{hex}}$} & \textbf{hit$@1$} & \textbf{$n_{\mathrm{seeds}}$} \\
\midrule
img & \texttt{LoRA-\allowbreak Noise} & 1000 & $+6e-06$ & $+1e-05$ & $1.00$ & 3 \\
img & \texttt{LoRA-\allowbreak Noise} & 3000 & $+1e-05$ & $+2e-05$ & $1.00$ & 3 \\
img & \texttt{LoRA-\allowbreak Noise} & 5000 & $+1e-05$ & $+2e-05$ & $1.00$ & 3 \\
img & \texttt{LoRA-\allowbreak Noise-\allowbreak SYNTH} & 1000 & $+8e-06$ & $+1e-05$ & $1.00$ & 3 \\
img & \texttt{LoRA-\allowbreak Noise-\allowbreak SYNTH} & 3000 & $+1e-05$ & $+2e-05$ & $1.00$ & 3 \\
img & \texttt{LoRA-\allowbreak Noise-\allowbreak SYNTH} & 5000 & $+9e-06$ & $+2e-05$ & $1.00$ & 3 \\
img & \texttt{bSFT-\allowbreak GUI} & 1000 & $+4e-05$ & $+6e-05$ & $1.00$ & 3 \\
img & \texttt{bSFT-\allowbreak GUI} & 3000 & $+0.0002$ & $+0.0003$ & $1.00$ & 3 \\
img & \texttt{bSFT-\allowbreak GUI} & 5000 & $+0.0004$ & $+0.0007$ & $1.00$ & 3 \\
img & \texttt{bSFT-\allowbreak GUI} & 7000 & $+0.0005$ & $+0.0008$ & $1.00$ & 3 \\
img & \texttt{bSFT-\allowbreak GUI} & 10000 & $+0.0015$ & $+0.0029$ & $1.00$ & 3 \\
img & \texttt{bSFT-\allowbreak Random} & 1000 & $+3e-05$ & $+5e-05$ & $1.00$ & 3 \\
img & \texttt{bSFT-\allowbreak Random} & 3000 & $+9e-06$ & $+2e-05$ & $1.00$ & 3 \\
img & \texttt{bSFT-\allowbreak Random} & 5000 & $+1e-05$ & $+2e-05$ & $1.00$ & 3 \\
img & \texttt{bSFT-\allowbreak SYNTH} & 1000 & $+0.0002$ & $+0.0001$ & $1.00$ & 3 \\
img & \texttt{bSFT-\allowbreak SYNTH} & 3000 & $+0.0003$ & $+0.0003$ & $1.00$ & 3 \\
img & \texttt{bSFT-\allowbreak SYNTH} & 5000 & $+0.0004$ & $+0.0003$ & $1.00$ & 3 \\
img & \texttt{bSFT-\allowbreak Safety} & 1000 & $+2e-05$ & $+4e-05$ & $1.00$ & 3 \\
img & \texttt{bSFT-\allowbreak Safety} & 3000 & $+2e-06$ & $+1e-05$ & $1.00$ & 3 \\
img & \texttt{bSFT-\allowbreak Safety} & 5000 & $+2e-06$ & $+1e-05$ & $1.00$ & 3 \\
img & \texttt{baseline} & 0 & $+0e+00$ & $+0e+00$ & $1.00$ & 3 \\
\midrule
txt & \texttt{T-\allowbreak LoRA-\allowbreak Noise-\allowbreak GUI} & 1000 & $+2e-06$ & $+4e-05$ & $1.00$ & 3 \\
txt & \texttt{T-\allowbreak LoRA-\allowbreak Noise-\allowbreak GUI} & 3000 & $+7e-06$ & $+3e-05$ & $1.00$ & 3 \\
txt & \texttt{T-\allowbreak LoRA-\allowbreak Noise-\allowbreak GUI} & 5000 & $+1e-05$ & $+8e-05$ & $1.00$ & 3 \\
txt & \texttt{T-\allowbreak LoRA-\allowbreak Noise-\allowbreak Safety} & 1000 & $+3e-06$ & $+2e-05$ & $1.00$ & 3 \\
txt & \texttt{T-\allowbreak LoRA-\allowbreak Noise-\allowbreak Safety} & 3000 & $+7e-06$ & $+4e-05$ & $1.00$ & 3 \\
txt & \texttt{T-\allowbreak LoRA-\allowbreak Noise-\allowbreak Safety} & 5000 & $+7e-06$ & $+3e-05$ & $1.00$ & 3 \\
txt & \texttt{T-\allowbreak bSFT-\allowbreak GUI} & 1000 & $+3e-05$ & $+0.0002$ & $1.00$ & 3 \\
txt & \texttt{T-\allowbreak bSFT-\allowbreak GUI} & 3000 & $+7e-05$ & $+0.0027$ & $1.00$ & 3 \\
txt & \texttt{T-\allowbreak bSFT-\allowbreak GUI} & 5000 & $+9e-05$ & $+0.0133$ & $0.88$ & 3 \\
txt & \texttt{T-\allowbreak bSFT-\allowbreak Random} & 1000 & $+2e-05$ & $+0.0002$ & $1.00$ & 3 \\
txt & \texttt{T-\allowbreak bSFT-\allowbreak Random} & 3000 & $+2e-05$ & $+0.0002$ & $1.00$ & 3 \\
txt & \texttt{T-\allowbreak bSFT-\allowbreak Random} & 5000 & $+3e-05$ & $+0.0002$ & $1.00$ & 3 \\
txt & \texttt{T-\allowbreak bSFT-\allowbreak SYNTH} & 1000 & $+0.0003$ & $+0.0007$ & $1.00$ & 3 \\
txt & \texttt{T-\allowbreak bSFT-\allowbreak SYNTH} & 3000 & $+0.0011$ & $+0.0012$ & $1.00$ & 3 \\
txt & \texttt{T-\allowbreak bSFT-\allowbreak SYNTH} & 5000 & $+0.0015$ & $+0.0013$ & $1.00$ & 3 \\
txt & \texttt{T-\allowbreak bSFT-\allowbreak Safety} & 1000 & $+0.0076$ & $+0.0003$ & $1.00$ & 3 \\
txt & \texttt{T-\allowbreak bSFT-\allowbreak Safety} & 3000 & $+0.0097$ & $+0.0004$ & $1.00$ & 3 \\
txt & \texttt{T-\allowbreak bSFT-\allowbreak Safety} & 5000 & $+0.0150$ & $+0.0004$ & $1.00$ & 3 \\
txt & \texttt{T-\allowbreak baseline} & 0 & $+0e+00$ & $+0e+00$ & $1.00$ & 3 \\
\midrule
U-img & \texttt{U-\allowbreak LoRA-\allowbreak Noise} & 1000 & $-0.0003$ & $-0.0011$ & $0.00$ & 3 \\
U-img & \texttt{U-\allowbreak LoRA-\allowbreak Noise} & 3000 & $-0.0001$ & $+0.0011$ & $0.00$ & 3 \\
U-img & \texttt{U-\allowbreak bSFT-\allowbreak GUI} & 1000 & $-0.0033$ & $-7e-05$ & $0.00$ & 3 \\
U-img & \texttt{U-\allowbreak bSFT-\allowbreak GUI} & 3000 & $-0.0070$ & $+0.0076$ & $0.00$ & 3 \\
U-img & \texttt{U-\allowbreak bSFT-\allowbreak Random} & 1000 & $+0.0038$ & $+0.0102$ & $0.00$ & 3 \\
U-img & \texttt{U-\allowbreak bSFT-\allowbreak Random} & 3000 & $+0.0034$ & $+0.0093$ & $0.00$ & 3 \\
U-img & \texttt{U-\allowbreak bSFT-\allowbreak SYNTH} & 1000 & $+0.0471$ & $+0.0260$ & $0.00$ & 3 \\
U-img & \texttt{U-\allowbreak bSFT-\allowbreak SYNTH} & 3000 & $+0.0544$ & $+0.0243$ & $0.00$ & 3 \\
U-img & \texttt{U-\allowbreak bSFT-\allowbreak Safety} & 1000 & $-0.0005$ & $-0.0006$ & $0.00$ & 3 \\
U-img & \texttt{U-\allowbreak bSFT-\allowbreak Safety} & 3000 & $-0.0001$ & $+0.0012$ & $0.00$ & 3 \\
U-img & \texttt{baseline} & 0 & $+0e+00$ & $+0e+00$ & $0.00$ & 3 \\
\bottomrule
\end{tabular}
\caption{All-cells systematic table. ``Mod.''\ = canary modality
(img = saturated image canaries, txt = text canaries, U-img = undertrained image canaries).
$\Delta$ values are means across $3$ seeds, paired-baseline subtraction per canary.
Compare against canonical cells C1--C5 in Table~\ref{tab:dissociation}.}
\label{tab:allcells}
\end{table}

Table~\ref{tab:allcells} reports $\Delta\mathrm{NLL}_{\mathrm{mean20}}$,
$\Delta\mathrm{NLL}_{\mathrm{hex}}$, and hit$@1$ for every
(canary modality, variant, step) cell in our matrix, averaged across
$3$ seeds. Per-cell hierarchical-bootstrap ($B\!=\!10000$) and
between-seed Student-$t_2$ $95\%$ CIs on
$\Delta\mathrm{NLL}_{\mathrm{mean20}}$ and $\Delta\mathrm{NLL}_{\mathrm{hex}}$
are computed for the $48$ CI-bearing rows ($34$ bSFT $+$ $14$
LoRA-Noise; baselines have $\Delta\!=\!0$ by construction); summary
counts for the headline cells are reported in App.~\ref{app:perseed}.

\paragraph{No descriptively clear-of-zero negative hex anywhere.}
Across all $48$ CI-bearing cells, \emph{zero} cells have a
hierarchical-bootstrap $95\%$ CI for $\Delta\mathrm{NLL}_{\mathrm{hex}}$
strictly below zero. Three cells have a slightly negative point
estimate (\textsc{U-LoRA-Noise} $-0.00113$, \textsc{U-bSFT-GUI}
$-7\!\times\!10^{-5}$, \textsc{U-bSFT-Safety} $-0.00062$; all
undertrained-regime step~$1000$, where baselines have hex NLL $2.6$
and hit$@1\!=\!0$); each brackets zero under both hier-bootstrap and
$t_2$ between-seed intervals.

\section{Cell accounting and matrix counts}
\label{app:scatter}

The matrix decomposes as $4$ bSFT variants
($\textsc{GUI}$, $\textsc{SYNTH}$, $\textsc{Safety}$, $\textsc{Random}$)
$\times\,3$ canary regimes (saturated image, text, undertrained image)
$\times\,3$ seeds $\times\,3$--$5$ step counts.

\begin{table}[H]
\centering
\small
\setlength{\tabcolsep}{3pt}
\begin{tabular}{@{}lrrcc@{}}
\toprule
\textbf{Bucket} & \textbf{Aggr.\ rows} & \textbf{Seed-level runs} & \textbf{In ``$34$'' denom.?} & \textbf{CIs computed?} \\
\midrule
bSFT (GUI/SYNTH/Safety/Random)         & $34$ & $102$ & yes & yes \\
LoRA-Noise norm-matched controls       & $14$ & $42$  & no  & yes \\
Baselines ($0$-step rows)              & $3$  & $9$   & no  & no \\
\midrule
\emph{CI-bearing rows}                 & $\mathit{48}\,(=\!34\!+\!14)$ & --- & --- & --- \\
\emph{Printed all-cells table rows}    & $\mathit{51}\,(=\!48\!+\!3)$  & --- & --- & --- \\
\emph{Total seed-level runs}           & --- & $144\,(=\!102\!+\!42)$ & --- & --- \\
\bottomrule
\end{tabular}
\caption{Explicit accounting. The $34$-cell denominator covers only
the GUI/SYNTH/Safety/Random bSFT regimes (image:$14$, text:$12$,
undertrained image:$8$); the CI-bearing pool adds $14$ LoRA-Noise
controls (baselines have $\Delta\!=\!0$); printed
Table~\ref{tab:allcells} additionally lists $3$ baseline rows.}
\label{tab:accounting}
\end{table}

\section{Merge-and-unload composition ablation}
\label{app:mergeunload}

\textbf{Stacked} probe = base (with merged canary) $+$ bSFT-LoRA loaded;
\textbf{Merged} probe = base (with merged canary) $+$ bSFT-LoRA merged
into the base, then unloaded. Canary memory is intact in both cases;
only the bSFT contribution changes from a LoRA delta to a full-rank
weight update. Both deltas use the same $M_{\mathrm{canary}}$-only
baseline. The ratio merged/stacked indicates how much of the
stacked-LoRA drift survives the LoRA-to-full-rank conversion.

The ``\textbf{Stacked}'' column in Table~\ref{tab:mergeunload} (C3
$+0.0126$, C5 $+0.0055$) comes from the merge-comparison aggregation
pipeline; the canonical $\Delta_{\mathrm{hex}}$ in
Table~\ref{tab:dissociation} (C3 $+0.0133$, C5 $+0.0076$) comes from
the $K\!=\!30$ token-level probe pass. The two paths agree in sign at
every cell; the magnitude gap ($\!\approx\!0.002$, max $0.0021$ on C5)
reflects merge-pipeline aggregation drift, not the bf16 path residual.
Table~\ref{tab:dissociation} remains the canonical headline estimator;
Table~\ref{tab:mergeunload} uses only stacked-vs-merged \emph{ratios}.

\textbf{Sanity check: merge does not erase all measured canary-span deltas.}
Secret-span deltas on C1, C3, C5 survive merging at ratios
$+0.98, +0.98, +0.77$; even on C4 itself the secret-span hex delta
survives at $+0.51$, so the bSFT delta is not wholesale erased.
Per-token NLL agreement (Table~\ref{tab:logit_agreement}) shows
mean $|\Delta\mathrm{NLL}_{\mathrm{stacked\_vs\_merged}}|\!\leq\!0.005$
on \texttt{canary\_hex} for C1--C4, while C4's \texttt{preamble}
$|\Delta|\!=\!0.02837$ is $\sim\!118\times$ its own secret-span
$|\Delta|\!=\!0.00024$. The stacked-vs-merged disagreement isolates to
C4's preamble tokens, consistent with the Case~2 claim that C4's
mean$_{20}$ does not reflect secret-span damage.

\begin{table}[H]
\centering
\scriptsize
\setlength{\tabcolsep}{4pt}
\begin{tabular}{@{}lrrrr@{}}
\toprule
\textbf{Cell} & \textbf{preamble} & \textbf{canary\_lit} & \textbf{canary\_hex} & \textbf{n tokens} \\
\midrule
C1 & $0.00002$ & $0.00000$ & $0.00100$ & $1800$ \\
C2 & $0.00041$ & $0.00002$ & $0.00011$ & $1800$ \\
C3 & $0.00003$ & $0.00000$ & $0.00388$ & $1800$ \\
C4 & $\mathbf{0.02837}$ & $0.00000$ & $0.00024$ & $1800$ \\
C5 & $0.00001$ & $0.00771$ & $0.04181$ & $1800$ \\
\bottomrule
\end{tabular}
\caption{Mean per-token $|\Delta\mathrm{NLL}|$ between stacked and
merged evaluation paths on the same canary captions
(${\sim}1800$ tokens per cell). C4's
preamble $|\Delta|$ is $\sim\!118\times$ its own \texttt{canary\_hex}
$|\Delta|$, matching the Mode~2 claim that C4's mean$_{20}$ sits on
\texttt{preamble} and is composition-sensitive there. C5's larger
\texttt{canary\_hex} number reflects bf16 sensitivity of the
undertrained baseline, not Mode~2.}
\label{tab:logit_agreement}
\end{table}

The C4 preamble collapse to $+0.05$ ratio is consistent with several
non-exclusive mechanisms (template-token-activation precision
sensitivity, BF16 merge rounding for low-magnitude weight updates,
LoRA-composition-specific effects) and we make no causal claim. The
weak conclusion --- a probe whose mass sits on \texttt{preamble} under
loaded-LoRA evaluation is not robust under merged-LoRA evaluation,
even though \texttt{canary\_hex} deltas on the same and other cells
are robust --- is sufficient to demote C4's mean$_{20}$ from a
memorization claim.

\begin{table}[H]
\centering
\scriptsize
\setlength{\tabcolsep}{3pt}
\begin{tabular}{@{}lcrrrrr@{}}
\toprule
\textbf{Cell} & \textbf{Span} & \textbf{Stacked} & \textbf{Merged} & \textbf{Merged/Stacked} \\
\midrule
C1 & preamble & $+0.0001$ & $+0.0001$ & +0.96 \\
   & canary\_hex & $+0.0029$ & $+0.0029$ & +0.98 \\
\addlinespace[2pt]
C2 & preamble & $+0.0004$ & $+3e-05$ & +0.07 \\
   & canary\_hex & $+0.0003$ & $+0.0002$ & +0.75 \\
\addlinespace[2pt]
C3 & preamble & $+6e-05$ & $+4e-05$ & +0.61 \\
   & canary\_hex & $+0.0126$ & $+0.0124$ & +0.98 \\
\addlinespace[2pt]
C4 & preamble & $+0.0297$ & $+0.0014$ & +0.05 \\
   & canary\_hex & $+0.0004$ & $+0.0002$ & +0.51 \\
\addlinespace[2pt]
C5 & preamble & $+2e-05$ & $+1e-05$ & +0.81 \\
   & canary\_hex & $+0.0055$ & $+0.0043$ & +0.77 \\
\bottomrule
\end{tabular}
\caption{Merge-and-unload composition ablation across the five canonical
cells. M$_{\mathrm{canary}}$ is merged into the base before bSFT in both
conditions; only the bSFT-LoRA composition differs. \textbf{Stacked}: bSFT
LoRA loaded as adapter on top of canary-merged base. \textbf{Merged}: bSFT
LoRA merged into the canary-merged base, then unloaded (no LoRA at probe
time). The C4 \textsc{T-bSFT-Safety} preamble drift collapses to near zero
under merging while C1, C3, C5 secret-span effects survive merging --- the
C4 signal is evaluation-stack-dependent (loaded vs.\ merged LoRA) (we do not claim a complete causal
mechanism).}
\label{tab:mergeunload}
\end{table}

\section{Direct re-exposure sanity check}
\label{app:posrecovery}

To check that the secret-span and behavioural probes \emph{can} move
in this testbed (mean20 was the only probe that moved on saturated
cells), we trained an additional T-bSFT-GUI variant where the GUI bSFT
data was \emph{augmented} with the $20$ full-template text-canary
captions themselves (full \texttt{TEXTCAN-\{16hex\}-END}, not a
$20$-token truncation) --- explicit re-exposure of the secret content,
not a benign-SFT condition. Across $3$ seeds:
$\Delta\mathrm{NLL}_{\mathrm{mean20}}\!=\!{+}4\!\times\!10^{-5}$ (vs.\
${+}9\!\times\!10^{-5}$ on plain T-bSFT-GUI; flat both ways),
$\Delta\mathrm{NLL}_{\mathrm{hex}}\!=\!{+}0.0006$ (vs.\ ${+}0.0133$;
$\sim\!95\%$ collapse), hit$@1\!=\!1.00$ (vs.\ $0.88$; recovers to
ceiling). The secret-span and behavioural probes move in the recovery
direction when secret content is explicitly re-trained on, while
mean20 stays flat. This is a probe-sensitivity sanity check, not a
Whack-a-Mole-style reactivation control: direct re-exposure is the
easiest possible positive contrast, and we use it only to address
``can the probes move?''.

\section{\texorpdfstring{C3 failure inventory and hit$@k$}{C3 failure inventory and hit@k}}
\label{app:c3failures}

For C3 (T-bSFT-GUI step $5000$ on text canaries) we report all greedy
generations across $20$ canaries $\times\,3$ seeds $=\,60$ generations
(hit$@1$) plus $k\!-\!1$ stochastic samples at $T\!=\!0.7,p\!=\!0.95$
for hit$@4$ and hit$@16$ (i.e.\ $3$ and $15$ samples beyond greedy, in
the convention defined in \S\ref{sec:method}).

\begin{figure}[H]
\centering
\includegraphics[width=0.85\columnwidth]{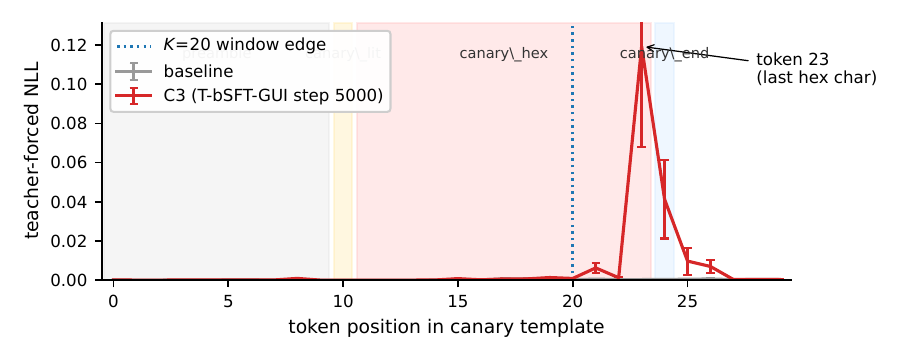}
\vspace{-3mm}
\caption{Position-aligned mean per-token NLL across $60$ (canary, seed)
pairs for C3 vs.\ baseline. Token $23$ (the final \texttt{canary\_hex}
BPE piece, containing the last $1$--$2$ hex characters; $1$ for
canary~$0$/$14$, $2$ for canary~$12$) jumps to $0.119$ ($\!\sim\!450\times$);
tokens $0$--$22$ flat. (Per-canary spikes vary; canary~$0$ alone moves
$+0.1228$ on its $13$-token hex span, hence canonical $+0.0133$ via
per-canary aggregation.)}
\label{fig:pertoken}
\end{figure}

\begin{table}[H]
\centering
\small
\setlength{\tabcolsep}{6pt}
\begin{tabular}{@{}lrrr@{}}
\toprule
\textbf{Seed} & \textbf{hit$@1$} & \textbf{hit$@4$} & \textbf{hit$@16$} \\
\midrule
seed 0 & $17/20$ & $18/20$ & $19/20$ \\
seed 1 & $18/20$ & $19/20$ & $19/20$ \\
seed 2 & $18/20$ & $19/20$ & $20/20$ \\
\midrule
total ($N\!=\!60$) & $53/60$ & $56/60$ & $58/60$ \\
\bottomrule
\end{tabular}
\caption{Per-seed hit$@k$ for C3. The damage is greedy-decode-specific:
hit$@1$ drops below $1.00$ on every seed, but the secret is recovered
under $k\!\geq\!4$ temperature samples on most or all canaries, indicating
the model still has the secret encoded but greedy decoding lands on a
near-miss tail.}
\label{tab:c3hitk}
\end{table}

All $7/60$ greedy failures are \emph{wrong hex tail} (last $1$--$2$
characters), e.g.\ target
\texttt{\seqsplit{TEXTCAN-139bed18a2190bba-END}} $\to$ greedy
\texttt{\seqsplit{TEXTCAN-139bed18a2190b68-END}}; this confirms the
Case-1 token-level analysis that damage concentrates on trailing hex
tokens outside $K\!=\!20$.

\paragraph{BPE token disclosure.}
Qwen2.5's BPE boundaries vary per canary, so the final
\texttt{canary\_hex} piece (token~$23$) holds the last $1$--$2$ hex
characters: $1$ for canaries~$0$/$14$ (which fail $3/3$ seeds), $2$
for canary~$12$ ($1/3$ seeds). Per-canary string, predicted-token,
and rank-of-correct-token manifests are authors' internal artifacts
and not released.

\paragraph{Leave-one-canary-out robustness.}
Headline $\Delta_{\mathrm{hex}}\!=\!{+}0.0133$. Removing canary~$0$
(its largest single-canary contributor, ${+}0.123$ on its own span)
gives ${+}0.0075$ ($\sim\!57\%$); the other $19$ leave-one-out point
estimates lie in $[{+}0.01332,{+}0.01400]$ ($<\!5\%$ of headline).
The behavioural break is concentrated in $3/20$ canaries with the
same Mode~1 token-$23$ pattern, not a single canary's idiosyncratic
tokenisation.

\section{Image--bSFT dataset overlap audit}
\label{app:overlap}

Image canaries draw from ScreenSpot~v2 \citep{wu2024osatlas}, a revised
release of the original ScreenSpot benchmark of \citet{cheng2024seeclick}, and the
same source as the \textsc{GUI} bSFT data; a near-duplicate would
trivially explain the C1 erosion as in-distribution relearning. We
verify no overlap two ways: \textbf{(i)} filename-set exclusion of the
$20$ canary images from bSFT sampling; \textbf{(ii)} perceptual-hash
check (\texttt{phash}, $8\!\times\!8$ DCT, distance $\le\!4$) across
all canary--bSFT image pairs --- no flagged pairs found in $20\times5000$
comparisons. The text-canary placeholder is a single $224\!\times\!224$
mid-gray bitmap shared by \textsc{Safety} and \textsc{Random} by
construction; a varied-image text-canary control is future work.

\section{Smaller-family decoy audit and Llama-1B trajectory details}
\label{app:crossfamily-decoy}

This appendix collects the smaller-family-only descriptive evidence
that the body summarises but does not print: the decoy table at the
candidate C5-equivalent cells (Tab.~\ref{tab:decoy}), the migrated
Qwen-1.5B C4-equivalent cell, and the Llama-3.2-1B step-trajectory
disagreements. None of this evidence pertains to the original 7B C5
cell, and we read all of it as descriptive smaller-family stress
tests (per the caption of Tab.~\ref{tab:cross-family}).

\begin{table}[H]
\centering\scriptsize
\setlength{\tabcolsep}{4pt}
\caption{\textbf{Decoy probe at the smaller-family candidate C5-equivalent cells (not the original 7B C5 cell).} For each model family, mean $\Delta$ relative to the \emph{true} target, averaged across $n=3$ seeds. If $\Delta_{\mathrm{mean20}}$ on shuffled-hex / wrong-secret / format-only decoys tracks the $\Delta$ on \texttt{true}, the in-window NLL drop is format/template smoothing rather than secret-specific learning. The original 7B C5 cell does not have a paired decoy audit and is not represented in this table.}
\label{tab:decoy}
\begin{tabular}{l l rrr}
\toprule
\textbf{Family} & \textbf{Decoy} & \textbf{$\Delta_{\mathrm{mean20}}^{\mathrm{vs.\,true}}$} & \textbf{$\Delta_{\mathrm{hex}}^{\mathrm{vs.\,true}}$} & \textbf{hit$@1$} \\
\midrule
Llama-3.2-1B & true & $+0.0000$ & $+0.0000$ & $0.000$ \\
Llama-3.2-1B & shuffled\_hex & $+4.1514$ & $+9.2485$ & $0.000$ \\
Llama-3.2-1B & wrong\_secret & $+2.5014$ & $+4.8954$ & $0.000$ \\
Llama-3.2-1B & format\_only & $+3.4520$ & $+10.8553$ & $0.000$ \\
\midrule
Qwen2.5-1.5B & true & $+0.0000$ & $+0.0000$ & $1.000$ \\
Qwen2.5-1.5B & shuffled\_hex & $+3.2050$ & $+6.4259$ & $0.000$ \\
Qwen2.5-1.5B & wrong\_secret & $+3.2336$ & $+4.3457$ & $0.000$ \\
Qwen2.5-1.5B & format\_only & $+1.6702$ & $+2.2722$ & $0.000$ \\
\bottomrule
\end{tabular}
\end{table}

\paragraph{Migrated Qwen-1.5B C4 cell.}
On Qwen2.5-1.5B the C4-equivalent cell migrates to T-bSFT-Random
($5000$): $\Delta_{\mathrm{mean20}}\!=\!{+}0.222$, $\Delta_{\mathrm{hex}}\!=\!{+}9\!\times\!10^{-4}$,
hit$@1\!=\!1.00$; $98\%$ of the mean drift sits on \texttt{preamble}
with \texttt{canary\_hex} essentially unchanged --- a sharper version
of the 7B C4 mechanism.

\paragraph{Llama-3.2-1B step-trajectory disagreements.}
Two phenomena absent from the 7B testbed appear at the Llama-1B
C3/C4-equivalent cells across step $\in\!\{1k,3k,5k\}$:
\textbf{(i)} on T-bSFT-GUI and T-bSFT-Safety the disagreement
\emph{grows} with bSFT magnitude rather than collapsing
($\Delta_{\mathrm{mean20}}\!=\!{+}0.56$ to ${+}0.81$ at step~$5000$
co-occurs with greedy hit$@1$ break $0.37$/$0.00$ while teacher-forced
\texttt{canary\_hex} stays within $1.6\!\times$ baseline);
\textbf{(ii)} T-bSFT-GUI shows emergent recovery in greedy hit$@16$
(seed-mean $0\!\to\!0.20\!\to\!0.78$) that is \emph{undetectable} via
$\mathrm{mean}_{K=20}$ NLL, which grows monotonically over the same
range. Inter-seed variance at step~$3000$ is high; we report a
descriptive trajectory, not a dose-response claim.

\end{document}